\documentclass[aps, prd, amsmath, floats, floatfix, twocolumn, nofootinbib, superscriptaddress, showpacs]{revtex4-1}
\usepackage{graphicx}
\usepackage{multirow}
\usepackage{color}

\usepackage{latexsym}
\usepackage{amsfonts}
\usepackage[colorlinks]{hyperref}
\graphicspath{{figures/}} 
\usepackage{tabularx}
\usepackage{gensymb}
\usepackage{bm}

\newcommand{\be}{\begin{eqnarray}}
\newcommand{\ee}{\end{eqnarray}}

\begin{document}

	\title{Phenomenological model for the electromagnetic response of a black hole binary immersed in magnetic field}

	\author{Carlos~A.~Benavides-Gallego}
	\email[corresponding author:]{cabenavidesg20@shao.ac.cn}
	\affiliation{Shanghai Astronomical Observatory, 80 Nandan Road, Shanghai 200030, P. R. China}

	\author{Wen-Biao Han}
	\email[corresponding author:]{wbhan@shao.ac.cn}
	\affiliation{Shanghai Astronomical Observatory, 80 Nandan Road, Shanghai 200030, P. R. China}
	\affiliation{School of Fundamental Physics and Mathematical Sciences, Hangzhou Institute for Advanced Study, UCAS, Hangzhou 310024, China}
			\affiliation{School of Astronomy and Space Science, University of Chinese Academy of Sciences, Beijing 100049, P. R. China}
			\affiliation{International Centre for Theoretical Physics Asia-Pacific, Beijing/Hangzhou 310024, China}

	\date{\today}

	\begin{abstract}
    Electromagnetic counterparts of gravitational wave events usually involve neutron stars during compact binary coalescences. On the other hand, the community generally believes that electromagnetic emissions are hardly generated during the mergers of binary black holes. Nevertheless, some authors have considered the possibility of an EM counterpart signal after the GW150914, opening the opportunity to investigate the electromagnetic emission of stellar-mass black holes (BHs) mergers. In the case of black holes in a strong magnetic field or with electric charges, electromagnetic emissions would appear accompanied by gravitational waves. In this work, based on the membrane paradigm, we consider a phenomenological model to study the electromagnetic radiation generated by binary black holes surrounded by a uniform magnetic field. We obtain the electromagnetic waveform detected by a far-distance observer for varied black hole spins. By analytical Fourier transformation, we get the chirp property of the electromagnetic waves; we found that the power-law is 5/6, which has the opposite sign compared to gravitational waves. The frequency of such electromagnetic waves is the same as gravitational waves.
	\end{abstract}

	\maketitle

		
	\section{Introduction}
	
	Gravitational waves (GW) were predicted by A. Einstein after the final formulation of the general theory of relativity (GR) in 1916~\cite{Einstein:1916cc, Einstein:1918btx} when he realized that the linearized weak-field version of the field equations has a waveform solution. Nevertheless, according to Einstein, the gravitational-wave amplitudes would be extraordinarily small to be detected. Moreover, scientists believed there was no astrophysical system that could generate gravitational radiation. Therefore, the physical reality of GW was in debate during its early stage.

	Then, in 1955, Josh Goldberg wrote a paper showing that binary star systems generate GW~\cite{Goldberg:1955zz}. In his work, using the Einstein-Infeld-Hoffman (EIH) approximation, J.~Goldberg found a contribution to the curvature tensor in the seventh order, in the equations of motion (up to ninth order), and radiation in the tenth order if one introduces the radiations terms in the 6th order of the transverse-transverse components. Furthermore, by calculating the energy-flux passing through a spherical surface at infinity, he found it was possible to determine the existence of gravitational radiation, which agrees with the definition of radiation in electromagnetism.

	The most significant moment in the development of gravitational-wave physics came in 1957, during the Chapel Hill conference\footnote{``\textit{Conference on the Role of Gravitation in Physics}''}. In that conference, thanks to the presentations of J.~Goldberg and Felix Pirani\footnote{ His contribution was called:``\textit{Measurement of
	Classical Gravitational Fields}''.}~\cite{Pirani:1956tn, Pirani:1956wr}, there was an intense discussion about the reality of gravitational waves and how to understand its interaction with a detector device in a physically meaningful way. In Pirani's contribution, for example, he proposed to connect the geodesic equation with Newton's second law by identifying the components of the Riemann tensor $R^{a}_{\;\;0b0}$ with the second derivative of the Newtonian potential. In this way, ``\textit{by measurements of the relative acceleration of several different pairs of particles, one may obtain full details about the Riemann tensor''}~\cite{Saulson:2010zz}. Although this idea did not mention gravitational waves, it would play a crucial role when applied to gravitational-wave physics because it showed that it was possible to construct a detector. In consequence, the physical reality of gravitational waves could be tested experimentally~\cite{Saulson:2010zz}. 

    After the Chapel Hill conference, Joseph Weber and John A.~Wheleer wrote a paper discussing the reality of gravitational waves. In their work, they showed that ``\textit{the disturbance in question is real and not removable by any change of coordinate system}'' \cite{Weber:1957oib}. Then, in 1959, J.~Weber published his work on the ``\textit{Detection and Generation of Gravitational Waves}” \cite{Weber:1960zz} where he proposed methods to measure the Riemann tensor, establishing in this way a program for building actual detectors. Hence, in 1969, Weber carried out the first attempt to construct a gravitational wave detector. His invention, known as Weber bars, consisted of several aluminum cylinders, 2m in length and 1m in diameter, connected to sensitive piezoelectric sensors capable of measuring changes in the longitude of cylinders by about $10^{-16}$m. In the same year, Weber wrote an article claiming the detection of gravitational radiation from the center of the Milky Way~\cite{Weber:1969bz}. Nevertheless, his results were not duplicated by others~\cite{Sciama:1974xj}. 

    Although the gravitational wave detector constructed by Weber was not good enough for detecting gravitational radiation, it opened the challenge for designing more sensitive devices. In 1972, for example, Rainer Weiss pictured a new kind of detector~\cite{Weiss:1972}. Instead of cylinders, Weiss proposed a Michelson interferometer to look for fluctuations in the Riemann tensor via phase-shift measurements, which is the most sensitive way of seeking such variations. Nevertheless, the technical difficulties in designing this kind of detector would take more than thirty years of labor. In the meantime, on the other hand, the discovery of the binary pulsar system PSR B1913-16 by Hulse and Taylor~\cite{Taylor:1979zz} and subsequent observations of its energy loss by Taylor and Weisberg~\cite{Taylor:1982zz} indirectly demonstrated the existence of gravitational waves.

    Nowadays, the construction of large laser interferometer detectors of gravitational waves (GWs), such as LIGO (US), VIRGO (Italy) and GEO 600 (Germany), TAMA300 (Japan), and the subsequent detection of the first GW signal on September 2015 by the Advanced LIGO~\cite{Abbott:2016blz}, has opened a new window of astrophysical research that would provide more tests of general relativity, especially in the strong-field regime~\cite{TheLIGOScientific:2016src, Abbott:2018lct}. From September 2015 to August 2017 (during the observation periods O1 and O2), the LIGO-Virgo collaboration has detected eleven GW events~\cite{LIGOScientific:2018mvr}. Some examples are: GW151226~\cite{Abbott:2016nmj}, GW170104~\cite{Abbott:2017vtc}, GW170814~\cite{Abbott:2017oio}, GW170817~\cite{Abbott:2017ntl,GBM:2017lvd,Monitor:2017mdv,ANTARES:2017bia,Abbott:2017wuw,Abbott:2017xzg}.  The majority of such events were associated with black hole-black hole (BHBH) mergers and one with a neutron star-neutron star (NSNS) merger. Recently, during the third observing run (O3) by LIGO and Virgo, the observation of GW from two compact objects with properties consistent with neutron star-black hole (NSBH) binaries has been announced in Ref.~\cite{LIGOScientific:2021qlt}. The first event, abbreviated as GW200105, was observed by LIGO Livingston and Virgo. While the second event, known as GW200115, was detected by all three LIGO–Virgo detectors. Furthermore, since the first gravitational wave detection (GW150914), there has been a huge effort to search GW events in previous observations~\cite{Abbott:2016ezn,Abbott:2016cjt,Abbott:2017ylp,Abbott:2017mnu,Abbott:2017pqa,Abbott:2017tlp,Abbott:2017cvf}, although no evidence has been found.

    The NSNS merger GW170817 was the first GW source followed by the detection of a gamma-ray burst (GRB)~\cite{Monitor:2017mdv}, a kilanova~\cite{Smartt:2017fuw,Arcavi:2017xiz,Pian:2017gtc}, and an off-axis GRB afterglow~\cite{Troja:2017nqp}. Thus, it has been considered as the first GW source accompanied by electromagnetic (EM) emission. Furthermore, the merger GW170817 has given new insights to several problems, such as the origin of short $\gamma$-ray bursts and the production of heavy elements~\cite{Levan:2020jlg}. In general, the community agrees that EM emissions are possible only during NSNS or NSBH mergers (see Ref.~\cite{Sylvestre:2003vc} for different mechanisms that may be important in generating EM emission). Nevertheless, some authors have considered the possibility of an EM counterpart signal after the GW150914, opening in this way the possibility to consider EM emission of stellar-mass black holes (BHs) mergers~\cite{Connaughton:2016umz}. Although it may be difficult to confirm that the GW150914 and the gamma-ray burst\footnote{Detected by the Fermi Gamma-ray Space Telescope $0.4$s after the merger.} are related, the EM observation has inspired several papers where the authors tried to connect EM emissions to stellar-mass black hole mergers~\cite{Perna:2016jqh, Li:2016iww, Zhang:2016rli, Loeb:2016fzn, Morsony:2016upv, Murase:2016etc, Kelly:2017xck,deMink:2017msu}. In Ref.~\cite{deMink:2017msu} for example, the authors propose a simple mechanism that requires the BH binary to have a circumbinary disk at the time of the merger. In this way, they can explain the EM signals following binary black hole mergers, with a delay of hours.

    Previous to the GW150914 event, some authors have considered the electromagnetic counterparts of BH binary mergers~\cite{Palenzuela:2009yr,Palenzuela:2009hx,Mosta:2009rr,Zanotti:2010xs}. In references~\cite{Palenzuela:2009yr, Palenzuela:2009hx} in particular, C.~Palenzuela et al. started to investigate the binary black holes' effects on the electromagnetic fields. To do so, the authors considered the dynamics of a black hole binary system during the merger showing that the plasma around the binary system can induce EM radiation: ``\textit{these time-dependent features will likely leave their imprint in processes generating detectable emissions and can be exploited in the detection of electromagnetic counterparts of gravitational waves}''~\cite{Palenzuela:2009yr}. Latter, in Ref.~\cite{Mosta:2009rr}, P.~Mosta et al. studied the vacuum electromagnetic counterparts of binary black hole mergers. In their work, the authors considered a simple model where the binary system moves in a uniform magnetic field anchored to a distant circumbinary disc~\cite{Mosta:2009rr}. To model the magnetic field, the authors take into account the Wald solution~\cite{Wald:1974np}. Hence, using astrophysically expected values for the electromagnetic fields, the authors solve the Einstein-Maxwell equations numerically. They found that the EM radiation in the lowest multipole expansion ($l=2$, $m=2$) accurately reflects the GW radiation. However, for higher $l$ modes, the amplitude evolution of the EM radiation is slightly large. The authors also found that the energy emission efficiency in the EM waves depends on the total spin, and its value is thirteen orders of magnitude smaller than the gravitational energy of realistic magnetic fields~\cite{Mosta:2009rr}. On the other hand, according to Ref.~\cite{Mosta:2009rr}, the EM radiation frequencies are outside those values that can be measured using radio observations. However, it is possible to detect them indirectly. For example:`` \textit{if the accretion rate of the circumbinary disc is small and sufficiently stable over the timescale of the final inspiral, then the EM emission may be observable indirectly as it will alter the accretion rate through the magnetic torques exerted by the distorted magnetic field lines}''~\cite{Mosta:2009rr}.

	Recently, L.~Liu et al. have studied the gravitational and electromagnetic radiation of BH binaries, with electric and magnetic charges~\cite{Liu:2020cds, Liu:2020vsy, Liu:2020bag}. In Ref~\cite{Liu:2020cds}, the authors considered first a primordial BH binary as point masses with charges moving in a Keplerian orbit. In this way, they calculated both the gravitational and electromagnetic radiation of the system, and the merger rate distribution, which is related to the parameter $\alpha$ (see Ref.~\cite{Liu:2020cds} and references therein for details). When the authors considered the extremal charged case, they found that $\alpha=12/11$, in contrast to uncharged primordial BH binaries where $\alpha=36/37$. Hence, the charge increases the merger rate. In Refs.~\cite{Liu:2020vsy, Liu:2020bag}, L.~Liu et al. studied the case of BH binaries with electric and magnetic chargers in circular and elliptical orbits on a cone. To do so, they consider a BH binary system formed by non-rotating dyonic black holes. Using a Newtonian method with radiation reactions, the authors calculated the total emission rate of energy and angular momentum generated by the gravitational and electromagnetic radiation. In the case of circular orbits, the authors showed that electric and magnetic charges significantly suppress the merger times of dyonic binaries~\cite{Liu:2020vsy}. On the other hand, when considering elliptical orbits, L.~Lui et al. showed that the emission rates of energy and angular momentum produced by the gravitational and EM radiation have the same dependence on the conic angle for different orbits~\cite{Liu:2020bag}. 

	In this work, we investigate the EM radiation of a BH binary system that is immersed in a uniform magnetic field and obtain the induced EM waveforms during the inspiral stage. Our study bases on a phenomenological model mentioned by C.~Palenzuela et al. in Ref~\cite{Palenzuela:2009hx}, which came as a consequence of  ``\textit{the membrane paradigm}'' \cite{Thorne:1986iy}. We organize the manuscript as follows. In Sec.~\ref{SecII}, we discuss the membrane paradigm. Then, in Sec.~\ref{SecIII}, we review the fundamental aspects in the motion of charged particles to compute the electromagnetic fields via the Liénard-Wiechert potentials. In Sec.~\ref{SecIV}, we obtain the electromagnetic wave, and we divide the discussion into four subsections. In Sec.~\ref{SecIVa}, we explain the effective-one-body (EOB) hamiltonian formalism for spinning black holes binaries. Hence, using the numerical data, we obtain the trajectory of the binary system and the GW. Then, using the results of Secs.~\ref{SecII} and ~\ref{SecIII}, we compute the EM waveform for the binary system in Sec.~\ref{SecIVb}. In Sec.~\ref{SecIVc}, we discuss the quasi-circular approximation to obtain an analytical expression for the Fourier transform of the EM signal in Sec.~\ref{SecIVd}. Finally, in Sec.~\ref{SecV}, we discuss our results. 
	
    In this manuscript, we denote vectors using bold letters and scalar with normal letters. On the other hand, we use $M$-units (see App.~\ref{A1}) in the plots. Nevertheless, since we follow Refs.~\cite{Landau:1975pou} and \cite{Maggiore:2007ulw}, we keep the gravitational constant $G$, the speed of light $c$ and electric constant $k_e$\footnote{Which is equal to unity in CGS units, see Ref.~\cite{Landau:1975pou}.} in all the expressions.
  
     \section{The membrane paradigm and the model \label{SecII}}
    
    The use of full GR in some astrophysical problems can be challenging. For this reason, it becomes necessary to include mental images, approximations, toy models, or paradigms to have a better understanding of certain phenomena. In black hole theory, this way of thinking has played a crucial role in facilitating new insights and subsequent development of the field. Hence,  ``\textit{the power of a paradigm is that it suggests what approximations are appropriate, or what features of the exact problem can be ignored in an analysis without losing the essence of the problem being studied}''~\cite{Thorne:1986iy}.  
     
    Before our current idea of a black hole, there were two viewpoints or paradigms. The ``\textit{frozen-star} and ``\textit{black hole}'' viewpoints~\cite{Thorne:1986iy,Thorne:1987bsa}. In the first case, viewed in Schwarzschild coordinates, the ``\textit{frozen-star}'' model considers a start that contracts rapidly. Then, the contraction slows down until it freezes at an infinite \textit{redshift surface}, the horizon. In the second case,  Oppenheimer and Snyder (using the Eddington-Finkelstein coordinates) showed that a star collapses to a singularity in a short time~\cite{Oppenheimer:1939ue}. Therefore, these paradigms are the result of different coordinate systems. Nevertheless, since the interior of the infinite redshift surface does not affect the external universe, the ``\textit{frozen-star}'' viewpoint prevailed over the ``\textit{black hole}'' point of view, and most of the theoretical work focused only on the first paradigm~\cite{Novikov:1965sik}.

    These paradigms agree that GR is the best description of gravity and, when applied correctly, both paradigms must give the same result. However, using Schwarzschild coordinates in the frozen-start model only helps to study physics outside the horizon. If one considers highly dynamic situations in which the horizon is essential, the frozen-start paradigm does not work, contrary to the black hole model. In Ref.~\cite{Thorne:1986iy}, the authors pointed out that most of the research work done using the frozen-star point of view does not include any relativistic effect. For example, the gravitational source is treated as a Newtonian monopole, and the horizon is an \textit{ad hoc} spherical surface placed near to GR prediction. In this sense, one can not model some black holes' problems using the frozen-star viewpoint because that approach produces ambiguous boundary conditions. One more example is the evolution of the magnetic dipole moment of a collapsing star. In the frozen star picture, the magnetic dipole moment reaches an asymptotic finite value as $t\rightarrow \infty$, while the correct answer is an outburst of radiation with a $1/t^5$ falloff, the proper calculation considers the dynamical nature of the space-time near the event horizon~\cite{DeLaCruz:1970kk, Price:1972pw}. 

    On the other hand, most of the astrophysical phenomena occurring far from the event horizon, but in which the event horizon has a crucial role, will generate problems if one uses the frozen star as a model~\cite{Thorne:1986iy}. One example is the process of electromagnetic extraction of energy~\cite{Blandford:1977ds}, where it is inadequate to replace the black hole with a surface of no return~\cite{Thorne:1986iy}. In this sense, it was necessary to develop a new paradigm to study black-hole physics without difficulty. A model that helps us to approximate the problems without losing the essence of the problem considered. This viewpoint is known as \textit{the membrane paradigm}.   
     
    The membrane viewpoint emerged due to some significant results obtained during the 1970s~\cite{Thorne:1986iy, Thorne:1987bsa}. First, we have the work of S.~W.~Hawking~\cite{Hawking:1974rv, Hawking:1974sw, Hawking:1976de} where he showed us how a stationary black hole radiates as a black body and follows the laws of thermodynamics if we relate the entropy of a  black hole with its surface area. Secondly, the discovery made by Hawking and Hartle in which entropy can be generated by deformation of the horizon due to external gravitational fields as if the horizon were viscous~\cite{Hawking:1972hy}. In the third place, the possibility to attribute an effective-charge density on the horizon, which can be polarized if one immerses a black hole in a static external electric field~\cite{Hanni:1973fn, Hajicek:1974oua}. Finally, the discovery in which the horizon behaves like a surface with electric resistivity when an electric current passes through a black hole~\cite{Blandford:1977ds}. Hence, taking into account these results, Damour and Znajek (independently) were able to express the horizon's evolution equations in such a way that it was possible to identify those terms similar to the electric conductivity, surface pressure, surface momentum, shear, and bulk viscosity, temperature, entropy, etc.~\cite{Damour:1978cg, Damour:1982ik}. However, because the Damour-Znajek approach treats the horizon as a 3-dimensional null surface embedded in 4-dimensional space-time with any relation to the external universe, the formalism was incomplete~\cite{Thorne:1986iy}. Therefore, to make Damour-Zanejk formalism a tool for black hole astrophysics, it was crucial to include a description of the external universe in which the horizon lives. This description is the ``$3+1$'' formulation of GR~\cite{Smarr:1979es, Piran:1982cy, Braginsky:1976rb, MacDonald:1982zz}, which considers the space-time as a family of surfaces of constant time (3-dimensional space-like hypersurfaces) and treat them as the usual 3-dimensional space evolving through a 1-dimensional time.
     
    The membrane paradigm is based on the ``$3+1$'' formulation of GR, and, from the mathematical point of view, it is equivalent to the black hole theory of GR. This paradigm considers the physics outside the event horizon, but particles and fields very near to the horizon have a highly complex, frozen, ``boundary-layer'' structure~\cite{Thorne:1986iy}. Hence, the membrane viewpoint ``\textit{stretches}'' the horizon to cover up the boundary layer to impose the membrane-like boundary conditions on the \textit{stretched horizon}. In other words, the null horizon is replaced with a time-like physical membrane endowed with electrical, mechanical, and thermodynamic properties, which is the essence of ``\textit{membrane paradigm}''. Nevertheless, It is important to point out that the membrane paradigm losses its validity inside the horizon. When an observer falls through the event horizon, he realizes that it does not have an electric charge and current. Only from outside the horizon seems to have these properties.
     
    The union of the 3+1 formulation of GR with the  Damour-Znajek membrane-horizon approach (\textit{the membrane paradigm}) has been a powerful tool for astrophysical studies, see, for example, Ref.~\cite{MacDonald:1982zz, Price:1986yy, Suen:1988kq, Zurek:1985gd, Frolov:1989jh}. In the particular case of GW radiation, we have the work of  C.~Palenzuela et al., where they studied the Binary black hole effects on electromagnetic fields~\cite{Palenzuela:2009hx, Palenzuela:2009yr}. In their work,  they consider a BHBH system surrounded by a circumbinary disk with a magnetic field. Then, by solving the Einstein-Maxwell equations (expressed in the $3+1$ formulation), they found the EM field evolves as follows. First, before the merger, the EM fields behave similarly to equals dipoles orbiting about each other. Then, close and through the merging phase, both the fields' strength and the EM energy flux increase. At this stage, the EM fields show a poloidal/toroidal magnetic/electric configuration, consistent with the  Blandford-Znajek mechanism. Finally, after the merger, the system behaves as Wald's solution~\cite{Wald:1974np}: a spinning sphere immersed in an external magnetic field. 
  	\begin{figure}[t]
  	\begin{center}
  	\includegraphics[scale=0.38]{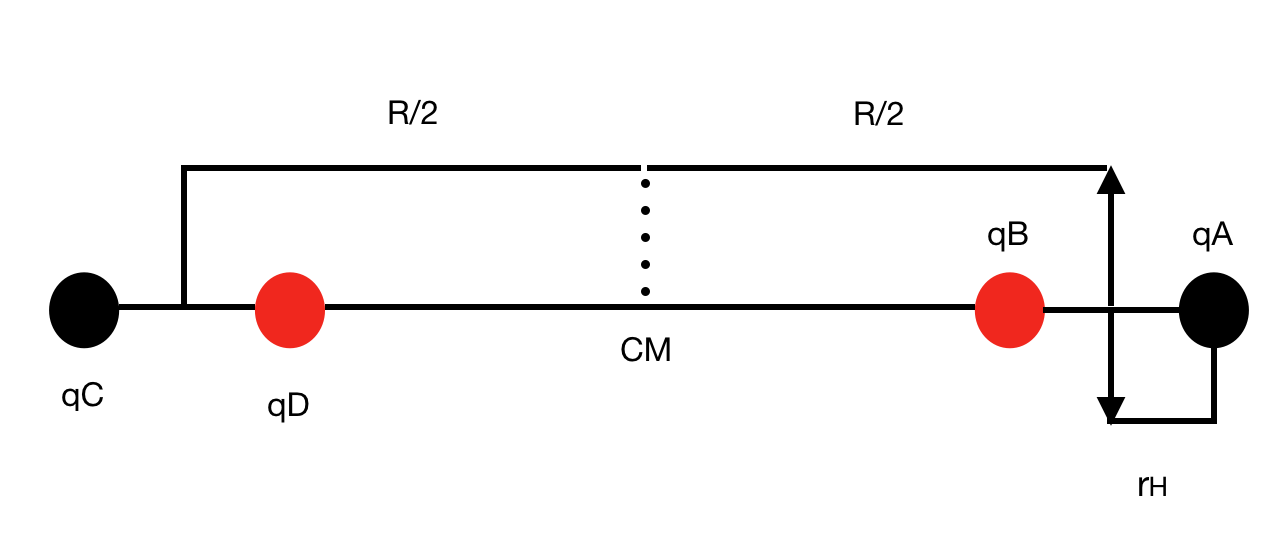}
  	\end{center}
  	\caption{Schematic representation of the phenomenological model based on the membrane paradigm. We use black ($q_A$ and $q_C$) and red ($q_B$ and $q_D$) colors to identify those charges with the same sign. \label{fig1}}
	\end{figure}
	
    To understand the phenomenological interaction between the fields and the binary system (before the merger), Palenzuela et al. use the membrane paradigm. First, the authors consider a surface charge density on the horizon of each black hole. Hence, due to the black holes' motion, the charge begins to separate perpendicularly to the velocity and magnetic field. This process induces an electric and magnetic field (different from the one anchored in the circumbinary disk). In this sense, as Palenzuela et al. pointed out, one can use a simple toy (phenomenological) model in which the black holes are replaced by four electric charges $q_i$ ($i=A, B, C, D$), whose values, when one assumes perfect conductivity of the horizon, are given by~\cite{Palenzuela:2009hx} 
     \begin{equation}
     \label{sec2e1}
     q_A=-\frac{r^2_H}{2\pi}\left|\frac{v\times B_0}{c}\right|=-q_B=q_C=-q_D.
     \end{equation}
    Here $v$ is the orbital velocity of the black holes, $B_0$ is the magnetic field anchored in the circumbinary disk, and $r_H$ is the radius of the apparent horizon. In Cartesian coordinates, the orbital motion of each charge is given by 
   	\begin{equation}
  	\label{sec2e2}
  	\textbf{W}_i=R_i(\cos\Omega t,\sin\Omega t,0),
  	\end{equation}
    where $i$ takes the values $A$, $B$, $C$. Hence, $R_A=R/2+r_H$, $R_B=R/2-r_H$, $R_C=-(R/2+r_H)$, and $R_D=-(R/2-r_H)$. $R$ is the distance between each black hole, see Fig.~\ref{fig1}. Using this phenomenological model, one can easily calculate the electromagnetic field induced by the BHBH binary system before the merger. Since the model involves electric charges in motion, it is necessary to use the Liénard-Wiechert potentials to obtain the correct form of the EM fields. In the next sections, we review and discuss the tools needed to get the electromagnetic waveform. In this sense, Sec.\ref{SecIII} is devoted to the Liénard-Wiechert potentials, which is the appropriated theoretical framework to calculate the EM fields of charges in motion. In Sec.\ref{SecIV}, we discuss the numeral method used to compute the black holes' trajectories, and we employ the data obtained from this simulation in the Liénard-Wiechert formulas.


	\section{The electromagnetic field of a system of charges at large distances \label{SecIII}}
	
    As discussed in the last section, the phenomenological model's main idea is to consider the BH binary system as a set of four charges. This idealization is a direct consequence of the separation process of the charge density (on the surface of the horizon) due to the black holes' motion. In this sense, the EM field generated by this set of charges should be computed by superposition, taking into account the movement of each charged particle.  
	
	\begin{figure}[t]
	\begin{center}
	\includegraphics[scale=0.38]{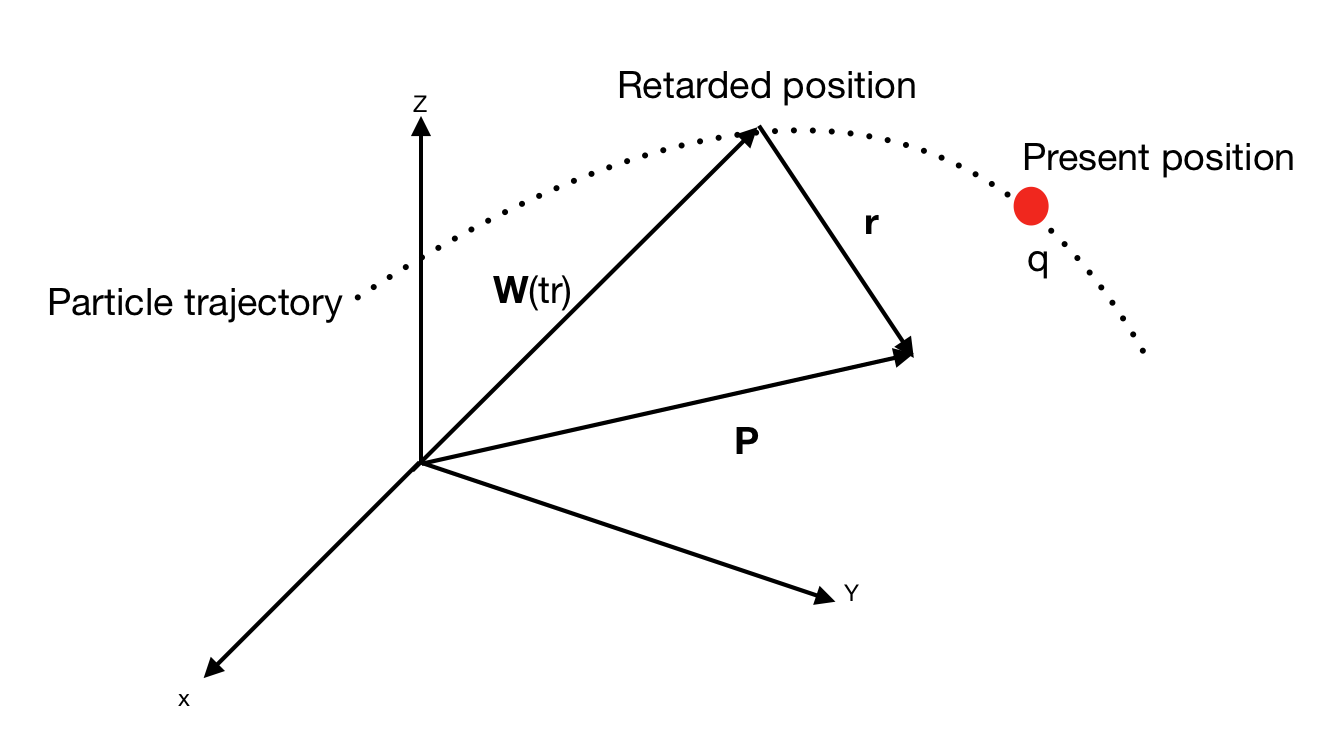}
	\end{center}
	\caption{The trajectory of a charged particle. The figure shows the retarded position $\textbf{W}(t_r)$ and the vector $\textbf{P}$, where the electric and magnetic fields are calculated. Note that the particles' position is evaluated at the retarded time $t_r$. \label{fig2}}
	\end{figure}
	
    When an electric charge is in motion, and one wants to compute the EM field in a particular point $\textbf{P}$, it is crucial to know that the field at that point has the retarded value. To understand this, let's consider one charged particle $q$ moving along a trajectory, see Fig.~\ref{fig2}. At each point, the value of the EM field is given by the position of $q$. Nevertheless, since no charged particle moves faster than the speed of light, we can assume that only the retarded point contributes to the fields at any given moment. Therefore, the fields must be evaluated at the retarded time $t_\text{r}$, which is determined implicitly by the relation~\cite{Landau:1975pou, Griffiths1981}
	\begin{equation}
	\label{3e1}
	||\mathbf{r}||=||\textbf{P}-\textbf{W}_r||=c(t-t_\text{r}),
	\end{equation} 
	 where $\textbf{W}_\text{r}=\textbf{W}(t_\text{r})$ is the vector position of the charge $q$ evaluated at the retarded time and $\textbf{r}$ is the vector from the retarded position to the point where an observed measures the EM field.
	 
	 According to electrodynamics\footnote{Here, we follow Ref.~\cite{Landau:1975pou} where $k_e=1$ (CGS units).}, the electric and magnetic fields are well defined by the retarded potentials~\cite{Landau:1975pou}
	 \begin{equation}
	     \label{3e2}
	     \begin{aligned}
	     \varphi(\textbf{r},t)&=\int\frac{\rho(\textbf{W}_r,t_r)}{||\textbf{r}||}dV,\\\\
	     \mathbf{A}(\textbf{r},t)&=\frac{1}{c}\int\frac{\mathbf{J}(\textbf{W}_r,t_r)}{||\textbf{r}||}dV.
	     \end{aligned}
	 \end{equation}
	 Here, $\rho(\textbf{W}_r,t_r)$ and $\textbf{J}(\textbf{W}_r,t_r)$ are the charge and current densities at the retarded time, respectively. Note that the potentials in Eq.~(\ref{3e2}), reduce to static case when $\rho$ and $\textbf{J}$ do not depend on time. 
	
	\begin{figure}[t]
	\begin{center}
	\includegraphics[scale=0.36]{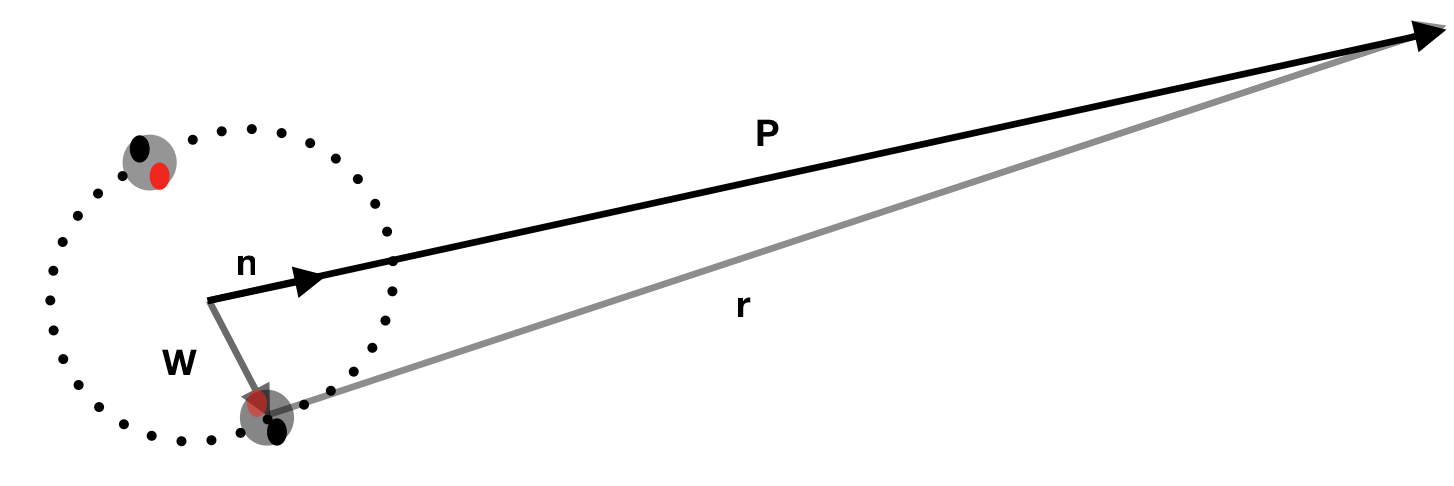}
	\end{center}
	\caption{Scheme of the binary system at large distances. In the figure, $\textbf{P}$ is the vector from the origin (center of mass) to the position at which the EM field is calculated. $\textbf{n}$ is a unit vector in the direction of $\textbf{P}$, and $\textbf{r}$ the vector from one of the charges to the point $\textbf{P}$.\label{fig2a}}
	\end{figure}
	 
	In the case of the EM field produced by a system of moving charges located at large distances in contrast to the dimension of the system, we can obtain an approximation for the retarded potentials of Eq.~(\ref{3e2}). To do so, we follow Ref.~\cite{Landau:1975pou}, where the origin of coordinates is taken anywhere in the interior of the system of charges, see Fig~\ref{fig2a}. In our case, we choose the origin of coordinates at the center of mass (CM) of the binary system. Hence, from the figure, we have that $\textbf{r}=\textbf{P}-\textbf{W}$. If we consider large distances, i. e. $||\textbf{P}||>>||\textbf{W}||$, we obtain~\cite{Landau:1975pou} 
	\begin{equation}
	    \label{3e3}
	    ||\textbf{r}||\approx||\textbf{P}||-\textbf{W}\cdot\textbf{n}.
	\end{equation}
    Under this approximation, the retarded time in Eq.~(\ref{3e1}) reduces to 
    \begin{equation}
        \label{3e4}
        t_r=t-\frac{||\textbf{P}||}{c}+\frac{\textbf{W}\cdot\textbf{n}}{c}
    \end{equation}
    and, the retarded potentials in Eq.~(\ref{3e2}), take the form~\cite{Landau:1975pou}
    \begin{equation}
        \label{3e5}
        \begin{aligned}
         \varphi(\textbf{r},t)&=\frac{1}{||\textbf{P}||}\int \rho(\textbf{W}_r,t_r)dV,\\\\
	     \mathbf{A}(\textbf{r},t)&=\frac{1}{c||\textbf{P}||}\int\mathbf{J}(\textbf{W}_r,t_r)dV.
	     \end{aligned}
    \end{equation}
    
    Before continuing our discussion, it is important to remark that the term $\textbf{W}\cdot\textbf{n}/c$ in Eq.~(\ref{3e4}) changes through time. Nevertheless, since we are considering the approximation at far distances, we can neglect this term and use $t_r=t-||\textbf{P}||/c$ as the retarded time. As we will see below, the contribution of the term $\textbf{W}\cdot\textbf{n}/c$ is already taken into account when we expand the retarded potentials in powers of $\textbf{W}\cdot\textbf{n}/c$.  
    
    At large distances, on the other hand, the EM field can be considered as a wave plane if one takes into account small regions of space\footnote{This requires the wavelength to be small compared with the distance at which the EM field is measured.}. As a consequence, it is possible to relate the electric and magnetic fields using the following relation~\cite{Landau:1975pou,Griffiths1981}
    \begin{equation}
        \label{3e6}
        \textbf{E}=\textbf{B}\times\textbf{n},
    \end{equation}
	which implies that $\textbf{E}$ and $\textbf{B}$ are perpendicular to each other. Hence, since $\textbf{B}=\nabla\times\textbf{A}$, it is clear that one only needs to compute the vector potential $\textbf{A}$ for a complete determination of the EM field in the \textit{wave zone}, which is the name of the region (located at large distances from the system of charges) where the wave plane approximation takes place. In this sense, expanding $\textbf{J}(\textbf{W}_r,t_r)$ in powers of $\textbf{W}\cdot\textbf{n}/c$ up to first order, one gets~\cite{Landau:1975pou}
    \begin{equation}
        \label{3e7}
        \textbf{A}=\frac{\dot{\textbf{d}}}{c||\textbf{P}||}+\frac{\ddot{\textbf{D}}}{6c^2||\textbf{P}||}+\frac{\bm{\dot{\mu}}\times\textbf{n}}{c||\textbf{P}||},
    \end{equation}
    where $\textbf{d}$ is the \textit{dipole moment} of the system defined as~\cite{Landau:1975pou}
    \begin{equation}
        \label{3e7}
            \textbf{d}=\sum_i q_i\textbf{W}_i,
    \end{equation}
    $\textbf{D}$ is the \textit{quadrupole moment} of the system with components\footnote{$D_{\alpha\beta}$ is the quadrupole moment tensor with null trace $D_{\alpha\alpha}=0$} $D_\alpha= D_{\alpha\beta}n_\beta$~\cite{Landau:1975pou}  
    \begin{equation}
        \label{3e8}
        D_{\alpha\beta}=\sum_iq_i(3x_\alpha x_\beta-\delta_{\alpha\beta}||\textbf{W}_i||^2),
    \end{equation}
    and $\bm{\mu}$ is the \textit{magnetic moment}, which is given by the relation~\cite{Landau:1975pou,Griffiths1981}
    \begin{equation}
    \label{3e9}
        \bm{\mu}=\frac{1}{2c}\sum_i q_i\textbf{W}_i\times \textbf{n}.
    \end{equation}
    In the last expressions, i. e. Eqs.~(\ref{3e7}), (\ref{3e8}) and (\ref{3e9}), the sum goes over all charges, the dot $\dot{{}}$ denotes derivative with respect to time and $x_\alpha$ are the components of $\textbf{W}$ for each charge. Hence, after computing $\nabla\times\textbf{A}$, we have that the EM field is given by the following expressions (see Ref.~\cite{Landau:1975pou} for details)
    \begin{equation}
    \label{3e10}
        \begin{aligned}
        \textbf{E}&=\frac{1}{c^2 ||\textbf{P}||}\left\{(\ddot{\textbf{d}}\times\hat{\textbf{n}})\times\textbf{n}+\frac{1}{6c}(\dddot{\textbf{D}}\times\textbf{n})\times\textbf{n}+\textbf{n}\times\dddot{\bm{\mu}}\right\},\\\\
        \textbf{B}&=\frac{1}{c^2 ||\textbf{P}||}\left\{\ddot{\textbf{d}}\times\textbf{n}+\frac{1}{6c}\dddot{\textbf{D}}\times\textbf{n}+(\dddot{\bm{\mu}}\times\textbf{n})\times\textbf{n}\right\},
        \end{aligned}
    \end{equation}
    where the expressions are evaluated at the retarded time $t_r$.
    
    From Eq.~(\ref{3e10}), we can see the contributions from the dipole (first term), quadrupole (second term), and magnetic (third term) moments to the EM field. Nevertheless, at far distances from the binary system, the dipole contribution vanishes, and the system is described by the quadrupole moment~\cite{Palenzuela:2009hx}. In this sense, we only use the quadrupole moment term to compute the EM wave, and Eq.~(\ref{3e10}) reduces to
    \begin{equation}
        \label{3e11}
            \begin{aligned}
            \textbf{E}&\approx \frac{(\dddot{\textbf{D}}\times\textbf{n})\times\textbf{n}}{6c^3||\textbf{P}||}\\
            \textbf{B}&\approx \frac{\dddot{\textbf{D}}\times\textbf{n}}{6c^3||\textbf{P}||}.
            \end{aligned}
    \end{equation}

	
	\section{The electromagnetic wave \label{SecIV}}
    
    The EM field produced by the system of charges in the phenomenological model (see Fig.~\ref{fig1}) requires obtaining the trajectory of the black hole binary system. To do so, we use a numerical simulation based on the effective-one-body approach (EOB). Then, from the numerical data, we use Eq.~(\ref{3e11}) to evaluate $\textbf{B}$ and $\textbf{E}$. Finally, we include the anchored magnetic field $B_0$ to the one produced by the system of charges. The final form of the fields at a given point involves implicit equations due to the different contributions depending on their respective retarded times.
	
	
	\subsection{Effective-one-body approach (EOB) \label{SecIVa}}
	
	Before 1999, analytical templates, based on the post-Newtonian (PN) approximation of the Einsteins fields equation, were developed to describe the inspiral stage of a binary sistem~\cite{Sasaki:2003xr, Blanchet:2006zz, Futamase:2007zz, Goldberger:2004jt}. However, these templates did not consider the last stages, such as the plunge, merger, and ringdown. Therefore, intending to consider these stages, A. Buonanno and T. Damour (1999) proposed a new approach to the two-body dynamics of compact objects. This new viewpoint is the so-called effective-one-body (EOB) approach~\cite{Buonanno:1998gg, Buonanno:2000ef}, later improved in Refs.~\cite{Damour:1999cr, Damour:2001tu,Buonanno:2005xu,Barausse:2009xi}. Currently, the EOB approach uses the PN theory, black-hole perturbation theory, and gravitational self-force formalism. The basic idea is to map the two-body problem onto an effective one-body problem through a canonical transformation. Therefore, the test particle moves in an effective external metric~\cite{Buonanno:1998gg}.
	
	\begin{figure*}[t]
		\begin{center}
		\includegraphics[scale=0.23]{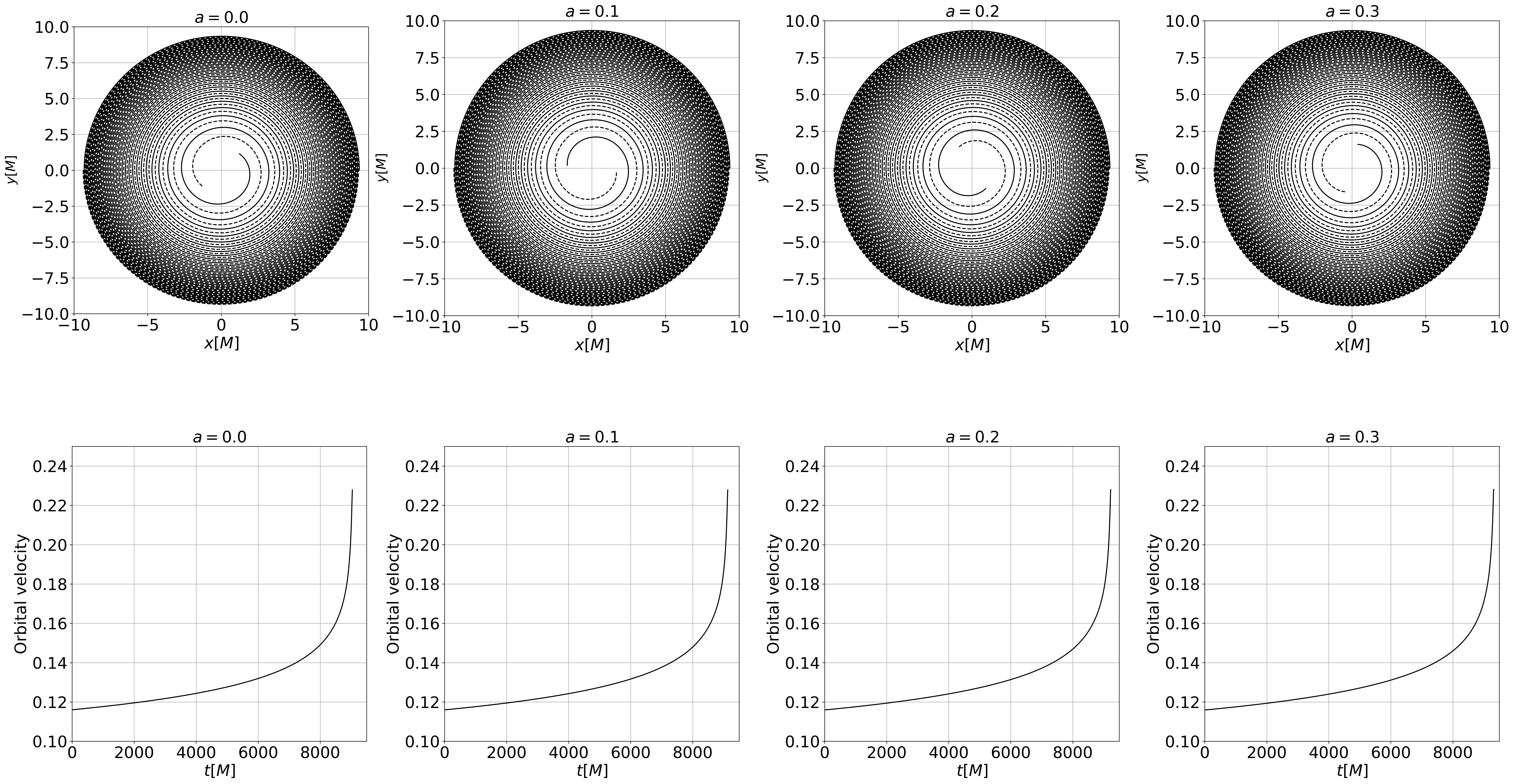}
		\includegraphics[scale=0.234]{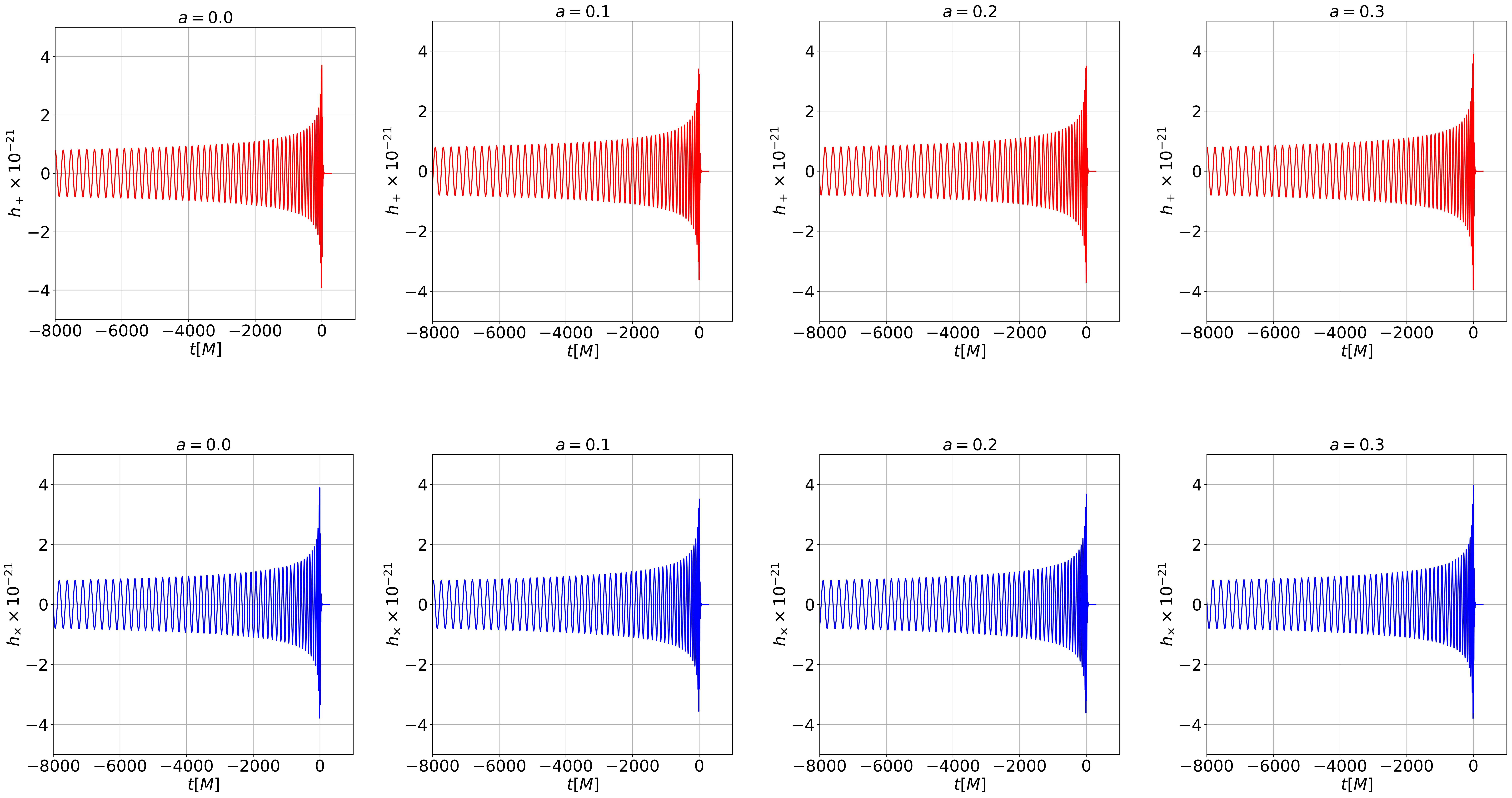}
		\end{center}
		\caption{Plots of the trajectory, orbital velocity, $h_+$ (red) and $h_\times$ (blue) obtained using the EOB approach. We consider $10M_{\odot}$ as the mass of each black hole. We use $M=20M_\odot$ units, the total mass of the system.\label{fig3}}
	\end{figure*}
	
    On the other hand, the numerical relativity (NR) simulations have helped to calibrate and improve the inspiral-merger-ringdown waveforms generated by the EOB~\cite{Pretorius:2004jg, Buonanno:2006ui}. Hence, the relation between NR and the EOB approach has been crucial for LIGO and VIRGO experiments. For example, during the search of high-mass and intermediate-mass black-hole binary in the data obtained by LIGO/Virgo colaboration~\cite{Abadie:2011kd}, the GW temples generated by the EOB model was used~\cite{Buonanno:2007pf}. In this sense, to get a suitable EM waveform from the phenomenological model, we use the EOB approach to obtain the most accurate trajectories for the black hole binary system.
	
	In Fig.~\ref{fig3}, we show the trajectory of each black hole in the binary system (first row) and the orbital velocity (which is the same for both black holes) obtained with the EOB approach, the second row in the figure. The trajectory of one of the black holes is given by the dashed line, while the continuous line describes the motion of the second black hole in the system. Therefore, it is easy to observe how the black holes approach each other. For the figure, we consider two cases: the Schwarzschild black hole (SBH) binary system and the Kerr black hole (KBH) binary system, where we take different values of the spin parameter: $a=0.1$, $0.2$ and $0.3$.  
	
	\begin{table*}
	\caption{Data from the numerical simulation using the EOB approach. The values correspond to the SBH binary system. The numerical solution gives us $t$, $R$, and $\phi$. The values for $\dot{R}$ and $\dot{\phi}$ are obtained by interpolation.\label{TableI}}
	\begin{ruledtabular}
	\begin{tabular}{ccccc}
	$t~[M]$ & $R~[M]$ & $\phi$ & $\dot{R}$& $\dot{\phi}~[M^{-1}]$\\
	\hline
	0.0000000000000000e+00&1.8744682583535340e+01&0.000000000000000e+00&-2.472898663775500e-04&1.2379714550577056e-02\\
	2.4783379715680933e+00&1.8743456847803685e+01&3.0681116647869152e-02&-2.472860512514998e-04&1.2380296055945360e-02\\
	4.9566759431361866e+00&1.8742231149892717e+01&6.1365115629408176e-02&-2.472697031559250e-04&1.2381532800842757e-02\\
	7.4350139147042800e+00&1.8741005494503003e+01&9.2052362420958059e-02&-2.472619221745544e-04&1.2382794635435204e-02\\
	9.9133518862723733e+00&1.8739779878314518e+01&1.2274261590766568e-01&-2.472542983094423e-04&1.2383995300525930e-02\\
	1.2391689857840467e+01&1.8738554300060411e+01&1.5343581400698653e-01&-2.472467324923855e-04&1.2385188088139614e-02\\
	1.4870027829408560e+01&1.8737328759449984e+01&1.8413197975536416e-01&-2.472392251143728e-04&1.2386390275337620e-02\\
	\end{tabular}
	\end{ruledtabular}
	\end{table*}
	
	In the Schwarzschild case ($a=0.0$), the initial separation between the two black holes is $R_0\approx18.744M$, and the initial value of the orbital velocity is $v_0\approx 0.1160$. In the case of a KBH binary system, the initial separation and the initial velocity are given by $R_0\approx18.738M$ $v_0\approx0.1159$, $R_0\approx18.732M$ $v_0\approx0.115944$, and $R_0\approx18.726M$ $v_0\approx0.115906$ when $a=0.1$, $0.2$ and $0.3$, respectively. On the other hand, from the second row of Fig.~\ref{fig3}, we see that the value of the orbital velocity increases as the each black hole approach to each other, reaching a maximum value. In the Schwarzschild case, $v_{max}\approx0.227814$. When we consider the KBH binary systme, we see that $v_{max}\approx0.227869$, $v_{max}\approx0.227956$, and  $v_{max}\approx0.228079$ for $a=0.1$, $0.2$, and $0.3$ respectively. Note how the binary system requires more time to reach the maximum value of the orbital velocity when the spin parameter $a$ increases.
	
	
	\subsection{Equations of motion and the electromagnetic wave\label{SecIVb}}

    In Sec.~\ref{SecII}, we mentioned that the motion of each charge in the phenomenological model is a circular orbit described by Eq.~(\ref{sec2e2}). Nevertheless, if one wants to consider a trajectory like those shown in Fig.~\ref{fig3}, it is necessary to point out that $R_i$ becomes a function of $t$ and $\Omega t$ changes to $\phi(t)$. Hence, the orbital motion (in cartesian coordinates) reduces to 
	\begin{equation}
	\label{4e1}
	\textbf{W}_i=\left\{R_i(t)\cos\phi(t),R_i(t)\sin\phi(t),0\right\},
	\end{equation}     	
    where the subscript $i$ takes the values A, B, C or D. According to Fig.~\ref{fig1}, we have two set of equations: one for charges A and B, and other for C and D. This set of equations are given by
	\begin{equation}
	\label{4e2}
	\begin{aligned}
	\textbf{W}_{A/B}&=\left\{\left(\frac{R(t)}{2}\pm r_H\right) 	 \cos\phi(t),\left(\frac{R(t)}{2}\pm r_H\right)\sin\phi(t)\right\},\\\\
	\textbf{W}_{C/D}&=-\textbf{W}_{A/B}.
	\end{aligned}
	\end{equation}
    Here the plus sing ($+$) is used for $\textbf{W}_A$, while the minus sing ($-$) for  $\textbf{W}_B$. On the other hand, the velocity $\textbf{v}_i$ and acceleration $\textbf{a}_i$ of each charged are given by
	\begin{equation}
	\label{4e3}
	\begin{aligned}
	\textbf{v}_i&=\frac{d\textbf{W}_i}{dt},\\
	\textbf{a}_i&=\frac{d\textbf{v}_i}{dt}=\frac{d^2\textbf{W}_i}{dt^2}.
	\end{aligned}	
	\end{equation}
    In Eq.~(\ref{4e2}), $R(t)$ is the separation between the black holes. Note that we are taking as the origin of coordinates the center of mass of the system. The values for $R(t)$ and $\phi(t)$ are given by the EOB approach, while $\dot{\phi}$ and $\dot{R}$ are computed by interpolation. Here the dot $\dot{}$ denotes derivative with respect to time $t$. In table~\ref{TableI}, we show some values for the binary system formed by two Schwarzschild black holes.
    
	Using Eq.~(\ref{4e2}), we can compute the EM field via Eq.~(\ref{3e11}). This is done by computing the vector $\textbf{D}$, which is the projection of the tensor $D_{\alpha\beta}$ along the unit vector $\textbf{n}$, see Eq.~(\ref{3e8}). Hence, if we consider $\textbf{n}$ to be the x-axis, i.  e.~$\{1,0,0\}$, we obtain
	\begin{equation}
	\label{4e4}
    \begin{aligned}
    D_x &= D_{xx}=\sum_i q_i(3x^2_i-||\textbf{W}_i||^2),\\
    D_y &= D_{yx}=\sum_i 3q_i y_ix_i,\\
    D_z &= D_{zx}=\sum_i 3q_iz_ix_i = 0   
    \end{aligned}
	\end{equation}
    Hence, we have that $\textbf{D}=\{D_x, D_y, 0\}$, $\dddot{\textbf{D}}=\{\dddot{D}_x,\dddot{D}_y,0\}$ and we obtain 
    \begin{equation}
    \label{4e5}
    \dddot{\textbf{D}}\times\hat{\textbf{n}}=
    \left|
    \begin{array}{ccc}
    \textbf{i}&\textbf{j}&\textbf{k}\\
    \dddot{D}_x&\dddot{D}_y&0\\
    1&0&0
    \end{array}
    \right|=\{0,0,-\dddot{D}_y\},
    \end{equation}
    from which
    \begin{equation}
    \label{4e6}
    (\dddot{\textbf{D}}\times\hat{\textbf{n}})\times \hat{\textbf{n}}=
    \left|
    \begin{array}{ccc}
    \textbf{i}&\textbf{j}&\textbf{k}\\
    0&0&-\dddot{D}_y\\
    1&0&0
    \end{array}
    \right|=\{0,-\dddot{D}_y,0\}.
    \end{equation}
    Therefore, at large distances (see Eq.~(\ref{3e11})) 
    \begin{equation}
    \label{4e7}
    \begin{aligned}
    \textbf{E}&\approx -\frac{1}{6c^3||\textbf{P}||}\dddot{D}_y \textbf{j}, \\\\
    \textbf{B}&\approx -\frac{1}{6c^3||\textbf{P}||}\dddot{D}_y \textbf{k}.
    \end{aligned}
    \end{equation}
    Now, using Eq.~(\ref{4e4}), $D_y$ has the form
    \begin{equation}
    \label{4e8}
    D_y= 3q_A y_A x_A+3q_B y_B x_B+3q_C y_C x_C+3q_D y_D x_D.
    \end{equation}
    Nevertheless, since $q_B=-q_A$, $q_C=q_A$, $q_D=-q_A$ and $y_C=-y_A$, $y_D=-y_B$, $x_C=-x_A$, and $x_D=-x_B$, the last expression takes the form
    \begin{equation}
    \label{4e9}
    \begin{aligned}
    D_y&=3q_A\left\{y_Ax_A-y_Bx_B+y_Ax_A-y_Bx_B\right\}\\
    &=6q_A\left\{y_Ax_A-y_Bx_B\right\}.
    \end{aligned}
    \end{equation}
    Hence, from the vectors $\textbf{W}_A$ and $\textbf{W}_B$, we obtain
    \begin{equation}
    \label{4e10}
    D_y=6q_A\left\{\left(\frac{R(t)}{2}+r_H\right)^2-\left(\frac{R(t)}{2}-r_H\right)^2\right\}\sin\phi(t)\cos\phi(t).
    \end{equation}
    Then, recalling that $2\sin\phi\cos\phi=\sin 2\phi$, the last expression reduces to
    \begin{equation}
    \label{4e11}
    D_y=3q_A r_H R(t)\sin 2\phi(t).
    \end{equation}
    Therefore,
    \begin{widetext}
    \begin{equation}
    \label{4e12}
    \begin{aligned}
    \dddot{D}_y&= 3r_H\big\{\sin 2\phi\left[3\dot{R}\left(\ddot{q}_A-4q_A \dot{\phi}^2\right)+3\dot{q}_A\ddot{R}+R\left(\dddot{q}_A-12\dot{\phi}\left(\dot{q}_A \dot{\phi}+q_A \ddot{\phi}\right)\right)+q_A \dddot{R}\right]\\\\
    &+2\cos2\phi\left[3R\ddot{q}_A \dot{\phi}+3 \dot{q}_A\left(2\dot{R}\dot{\phi}+R\ddot{\phi}\right)+q_A\left(3 \ddot{R} \dot{\phi}+3\dot{R}\ddot{\phi}+R\left(\dddot{\phi}-4\dot{\phi}^3\right)\right)\right]\big\}.
    \end{aligned}
    \end{equation}
    \end{widetext}
    
\begin{figure}[t]
		\begin{center}
		\includegraphics[scale=0.4]{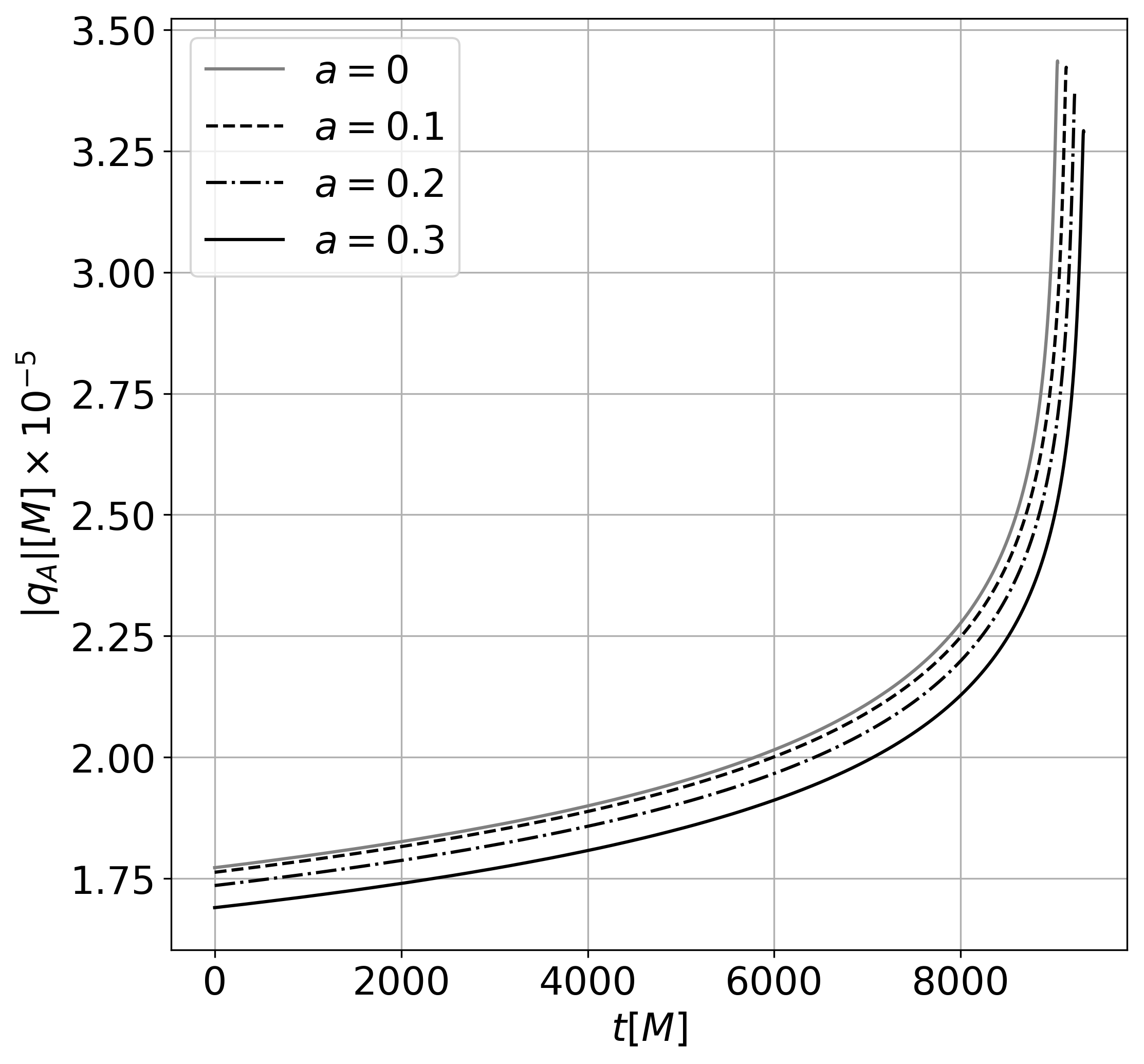}
		\end{center}
    \caption{Behavior of $|q_A|$ as a function of $t$ for different values of $a$. The plot is done using $M=20M_\odot$ units.\label{fig5}}
	\end{figure}

    From the astrophysical point of view, it is important to consider realistic values of the magnetic field anchored in the circumbinary disk with energy several orders smaller than the gravitational energy~\cite{Palenzuela:2009hx}. In this way, we guarantee that the effects of the electromagnetic fields on the geometry and black holes' dynamics are negligible. According to Palenzuela et al., one can use the value $B_0\simeq10^{16}(M_{\odot}/M)\text{Gauss}$ as the magnetic field on the circumbinary disk. We can express this value in units of $M^{-1}$ using the following relation~\cite{Palenzuela:2009hx}
	\begin{equation}
	\label{4e12}
	B[M^{-1}]=1.2\times 10^{-20}\left(\frac{M}{M_\odot}\right)B[\text{Gauss}].
	\end{equation}
    With the help of this consideration, we use Eq.~(\ref{sec2e1}) and the orbital velocity of the BH binary system to obtain $q_A$ as a function of time, see Fig.~\ref{fig5}. From the figure, as expected, we can see how the charge increases as the separation process on the surface of each black hole take place. However, during the inspiral phase, the increment of $q_A$ is slower as the spin parameter increases, and we can neglect terms with $\dddot{q}_A$ in Eq.~(\ref{4e11}).
    
    On the other hand, we must consider realistic astrophysical distances to locate an observer. According to the data, the merger of a binary system formed by two black holes produced the GW150914 event. This collision occurred at a distance of more than one billion \textit{light-years} (ly) \footnote{https://www.ligo.org/science/Publication-GW150914/}. In this sense, we assume the observer is located along the x-axis at a distance $L=||\textbf{P}||=1\times10^9\text{ly}$, which is equivalent to $L\approx3.2\times 10^{20}M$, see App.~\ref{A1}. 
    
	\begin{figure*}[t]
		\begin{center}
		\includegraphics[scale=0.235]{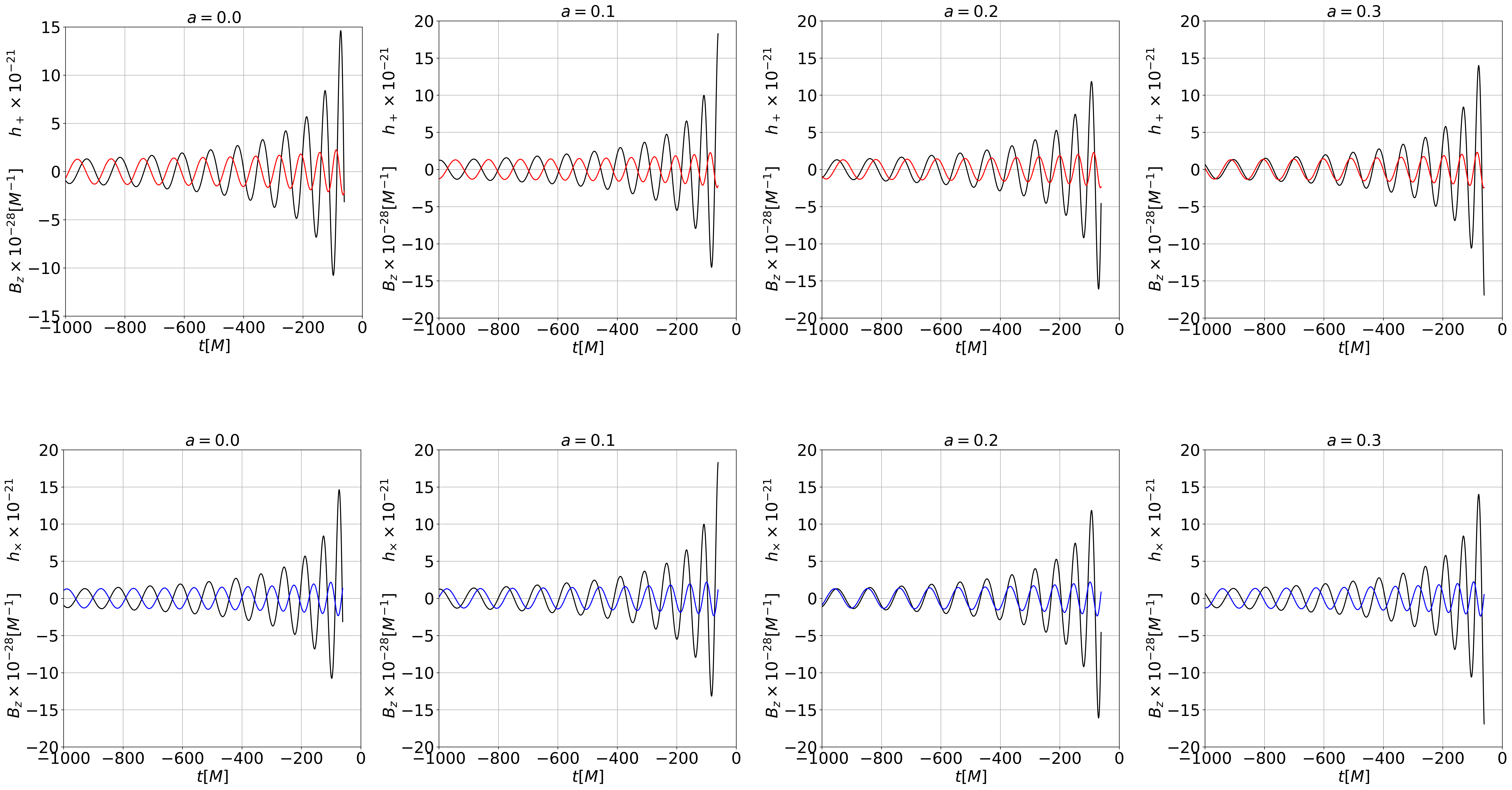}
		\end{center}
		\caption{Plots of $B_z$ (black), $h_+$ (red) and $h_\times$ (blue). We consider $10M_{\odot}$ as the mass of each black hole and $||\textbf{P}||\approx3.2\times 10^{20}M$. The plots are done using $M=20M_\odot$ units.\label{fig6}}
	\end{figure*}
    
	To understand the process of computing the EM wave, let's consider the motion of one of the charges. At $t_0$, the charge $q_A$ emits the first signal, which travels at the speed of light $c$ towards the observer, located at a distance $||\textbf{P}||$. The signal emission continues while the charge moves with a difference in time $\Delta t$. At $||\textbf{P}||/c$, the observer detects the first signal. Hence, the EM field measured by the observer is the one generated by the charge at $t_0$. Then, at $\Delta t+||\textbf{P}||/c$, the second signal arrives, and the observer measures the EM field generated by the charge at $t_0+\Delta t$, when the $q_A$ was at  $\textbf{W}_{A}(t_0+\Delta t)$. The process continues in the same way until some time before the black holes merge.
	
    The numerical data gives us the position of each charge, from which we compute the EM field using Eq.~(\ref{4e7}) evaluated at the retarded time $t_r$. Since the observer receives the first signal at $||\textbf{P}||/c$, the values of $\textbf{E}$ and $\textbf{B}$ correspond to the values of the EM field at the initial configuration of the system of charges. Therefore, at any $(t_k)_\text{Observer}=(t_k)_\text{Data}+||\textbf{P}||/c$, we use the values of $\textbf{E}$ and $\textbf{B}$ at $(t_k)_\text{Data}$. However, because $\Delta t_\text{Observer}=\Delta t_\text{Data}$, we can reset $t_\text{Observer}=0$ at the moment the observer receives the first signal, and plot the EM field using $t_\text{Data}$. The results are shown in Fig.~\ref{fig6}.
	
	\begin{figure*}[t]
		\begin{center}
		\includegraphics[scale=0.235]{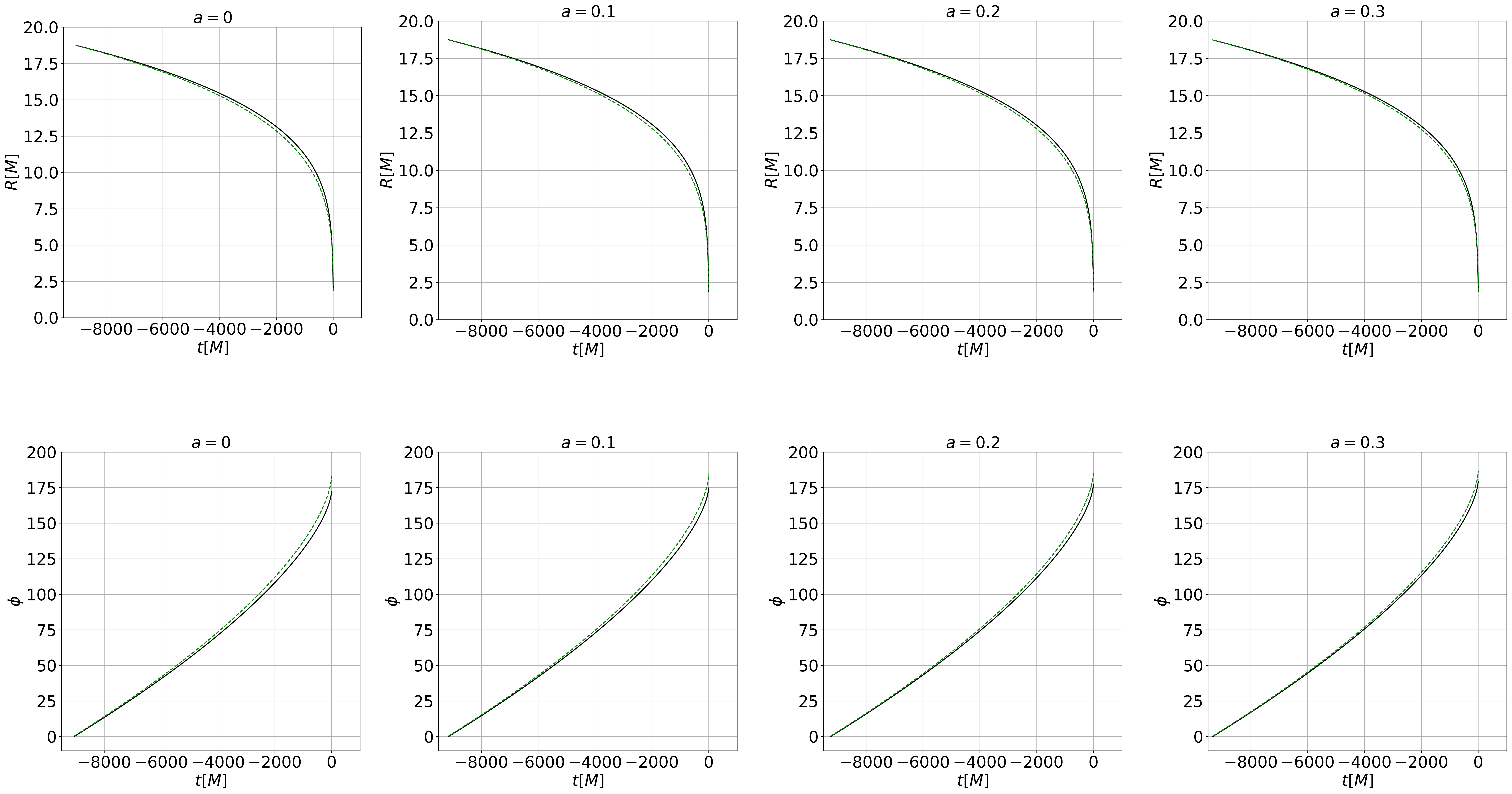}
		\end{center}
		\caption{Plots of $R$ and $\phi$ as functions of time $t$. In the figure the values given by the EOB approach are shown in black, while those in green-dashed line correspond to the quasi-circular approximation, see Eqs.~(\ref{4Ce12}) and (\ref{4Ce13}). The plots are done using $M=20M_\odot$ units.\label{fig7}}
	\end{figure*}
	
    
    \subsection{The quasi-circular approximation\label{SecIVc}}
    
    In Ref.~\cite{Maggiore:2007ulw}, M. Maggiore studies the inspiral of binary systems formed by compact objects such as stars, neutron stars, or black holes. Treating these objects as point-like particles moving on a Newtonian trajectory, Maggiore obtains some analytical results which help us to understand, in a simple way, the essence of the physical mechanisms, establishing the basic theoretical frame for more realistic astrophysical events. For this reason, in this subsection, we follow the same ideas, and we apply them to the phenomenological model presented in Secs.\ref{SecII} and \ref{SecIII}.
    
    In the Newtonian approximation, and considering the center of mass as the origin of coordinates, it is well-known that the dynamics of a binary system reduces to a one-body problem with \textit{reduced mass}
    \begin{equation}
        \label{4Ce1}
        \mu=\frac{m_1m_2}{m_1+m_2},
    \end{equation}
    where $m_1$ and $m_2$ are the mass of each compact object in the system. In our case, we always consider two black holes with $m_1=m_2=10M_\odot$. On the other hand, if the system moves in a circular Keplerian orbit, it is known that the total radiated power $\mathcal{P}$ emitted by the system is given by~\cite{Maggiore:2007ulw}
    \begin{equation}
        \label{4Ce2}
        \mathcal{P}=\frac{32}{5}\frac{c^5}{G}\left(\frac{GM_c\omega_{\text{GW}}}{2c^3}\right)^\frac{10}{3}. 
    \end{equation}
    Here $\omega_{\text{GW}}=2\omega_s$ is the angular velocity of the GW and $\omega_s$ the orbital frequency. $M_c$ is a quantity known as the \textit{chirp mass} and it is given by the relation~\cite{Maggiore:2007ulw} 
    \begin{equation}
        \label{4Ce3}
        M_c=\mu^\frac{3}{5}(m_1+m_2)^\frac{2}{5}=\frac{(m_1m_2)^\frac{3}{5}}{(m_1+m_2)^\frac{1}{5}}.
    \end{equation}
    Nevertheless, in a real astrophysical scenario, the radiation of GWs requires energy. In this sense, as the binary system emits gravitational radiation, the energy of the orbit, given by~\cite{Maggiore:2007ulw}
    \begin{equation}
        \label{4Ce4}
        E_{\text{orbit}}=E_{\text{kin}}+E_{\text{pot}}=-\frac{Gm_1m_2}{2R},
    \end{equation}
    must decreases as the system evolves. Therefore, $E_\text{orbit}$ becomes more and more negative and, as a consequence, the separation between the compact objects $R$ decreases in time. Thus, from the Kepler's law~\cite{Maggiore:2007ulw} 
    \begin{equation}
        \label{4Ce5}
        \omega^2_s=\frac{Gm}{R^3},
    \end{equation}
    with $m=m_1+m_2$ the total mass of the system, it is clear that the orbital frequency of the system and the radiated power increase as $R$ decreases. In this sense, the emission of GWs eventually leads to the coalescence of the binary system, see Fig.~\ref{fig3}. 

    In a realistic binary system, during the inspiral phase, the orbital frequency is small, and one can consider some approximations to obtain analytical expressions which describe the physics behind the interaction between the compact objects in the binary system. According to  Eq.~(\ref{4Ce5}), after solving for $R$ and computing the derivative with respect to time, the radial velocity is given by 
    \begin{equation}
        \label{4Ce6}
        \dot{R}=-\frac{2}{3}(\omega_s R)\frac{\dot{\omega_s}}{\omega^2_s}.
    \end{equation}
    From the last expression, we can conclude that $|\dot{R}|$ is much smaller than the tangential velocity $\omega_sR$ if
    \begin{equation}
        \label{4Ce7}
        \dot{\omega}_s<<\omega_s.
    \end{equation}
    In this way, as long as Eq.~(\ref{4Ce7}) holds, we can approximate the motion of the binary system as a circular orbit with a slow variation of $R$. This condition is known as the \textit{quasi-circular} approximation, and it is helpful to understand the evolution of the binary system during the inspiral phase. 

    In the case of a circular orbit, the energy is given by~\cite{Maggiore:2007ulw}
    \begin{equation}
       \label{4Ce8}
       E_\text{orbit}=-\left(\frac{G^2M^5_c\omega^2_\text{GW}}{32}\right)^\frac{1}{3}.
    \end{equation}
    However, because the emission of GWs requires energy, the conservation of energy\footnote{It is important to point out that the masses $m_1$ and $m_2$ do not have an internal structure. Therefore, the only possible source of energy is $E_\text{orbit}$. See Ref.~\cite{Maggiore:2007ulw} for details.} suggest that the radiated power $\mathcal{P}$ is equal to the change in $E_\text{orbit}$. Mathematically, we have that
    \begin{equation}
      \label{4Ce9}
      \mathcal{P}=-\frac{dE_\text{orbit}}{dt},
    \end{equation}
    from which, since $\omega_\text{GW}=2\pi f_\text{GW}$, one obtains~\cite{Maggiore:2007ulw}
    \begin{equation}
      \label{4Ce10}
      f_\text{GW}(\tau)=\frac{1}{\pi}\left(\frac{5}{256}\frac{1}{\tau}\right)^\frac{3}{8}\left(\frac{GM_c}{c^3}\right)^{-\frac{5}{8}}.
    \end{equation}
    In the last expression, $\tau$ represents the time taken by a GW wave to propagate from the source to the observer. Here, it is important to remark that $f_\text{GW}$ must be evaluated at the retarded time. Nevertheless, since the retarded time and the observer time differ only by a constant (i. e. $||\textbf{P}||/c$), we have that $\tau$ is given by
    \begin{equation}
        \label{4Ce11}
        \tau=(t_\text{coal})_\text{r}-t_\text{r}=t_\text{coal}-t,
    \end{equation}
    where $(t_\text{coal})_\text{r}$ denotes the value of the retarded time $t_\text{r}$ at the coalescence, and $t_\text{coal}$ is the value of the observer time $t$ at the coalescence, see Ref.~\cite{Maggiore:2007ulw}. 
	\begin{figure*}[t]
		\begin{center}$
			\begin{array}{cc}
			\includegraphics[scale=0.36]{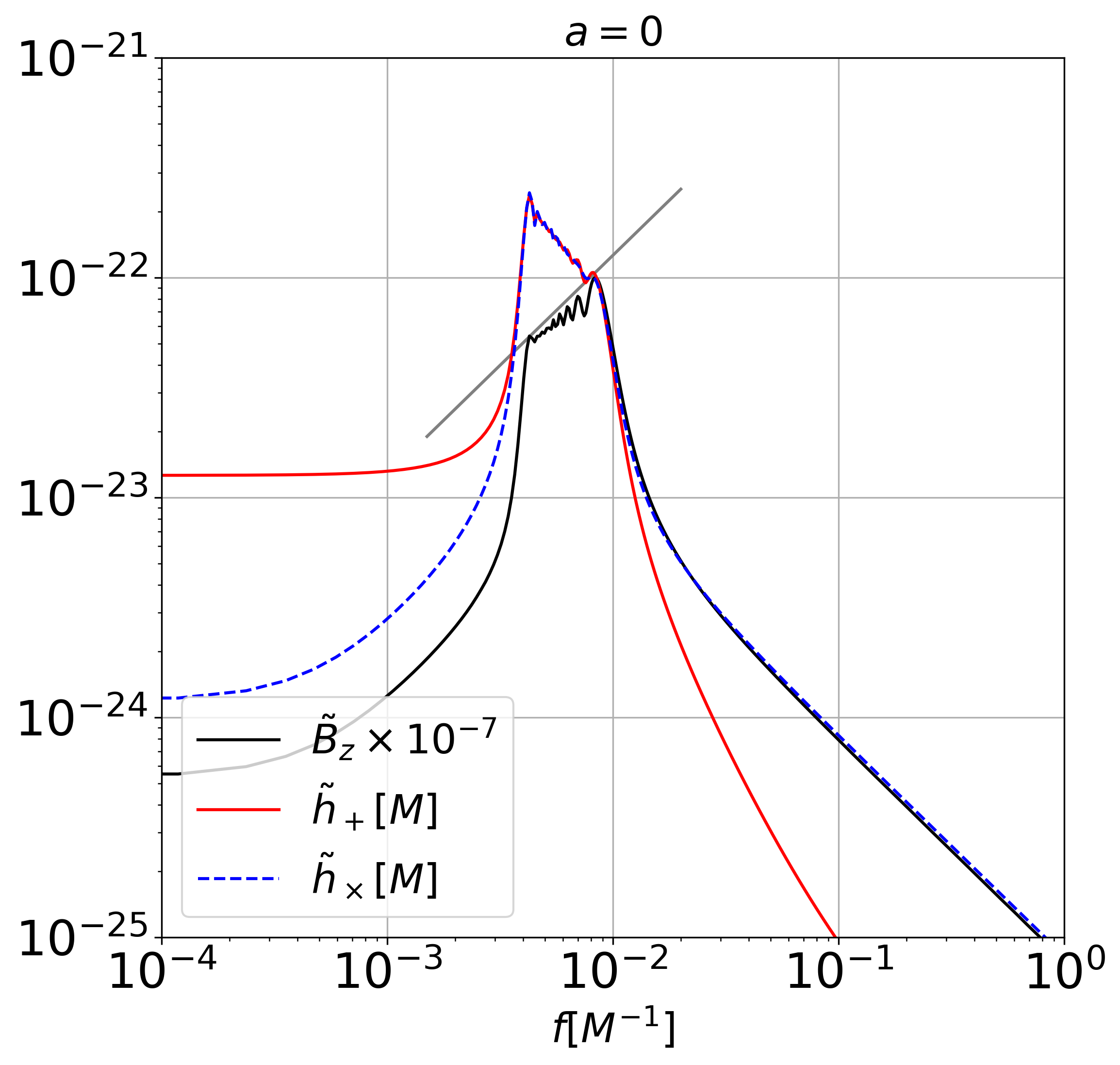}&
			\includegraphics[scale=0.36]{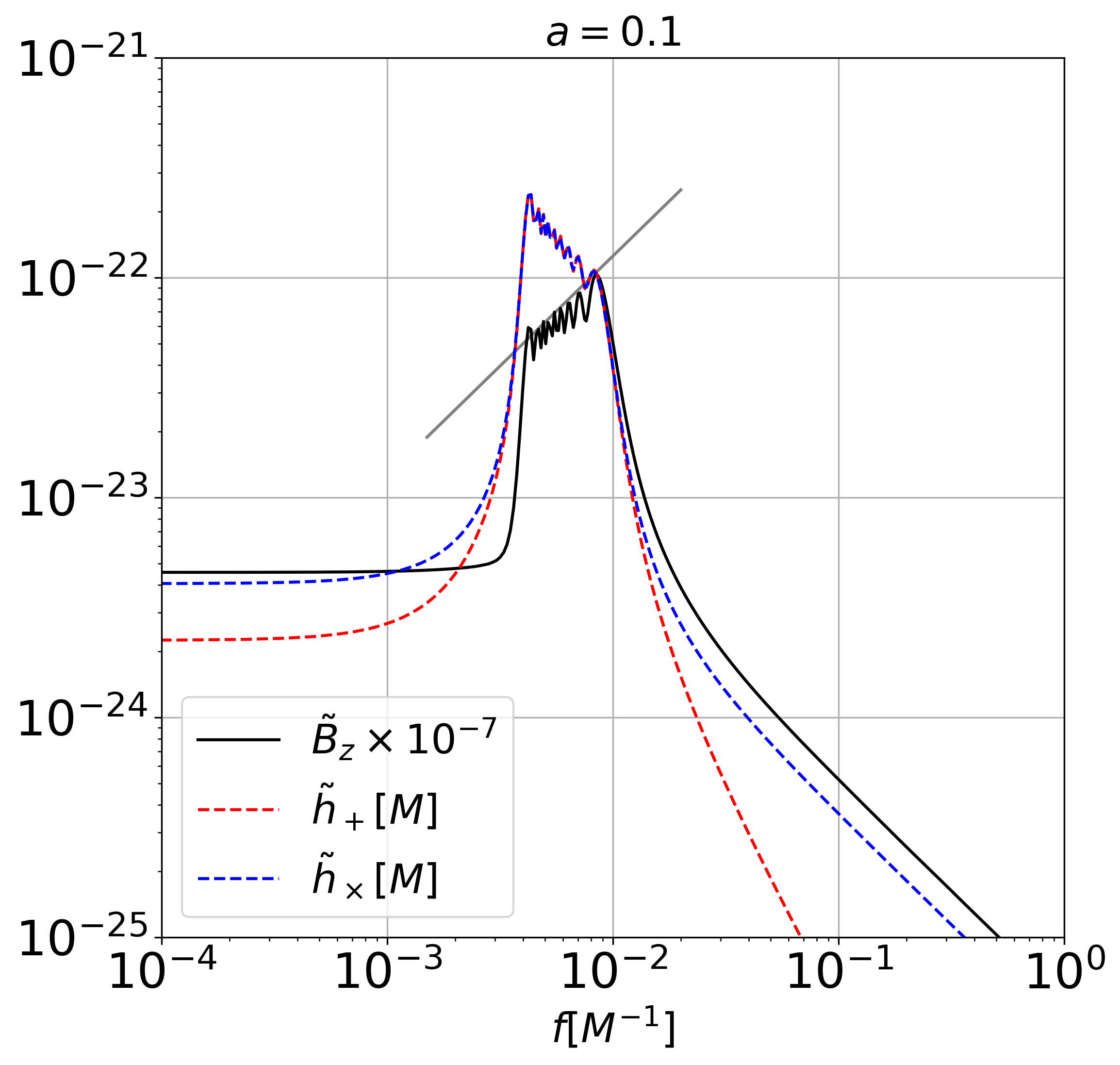}\\
			\includegraphics[scale=0.36]{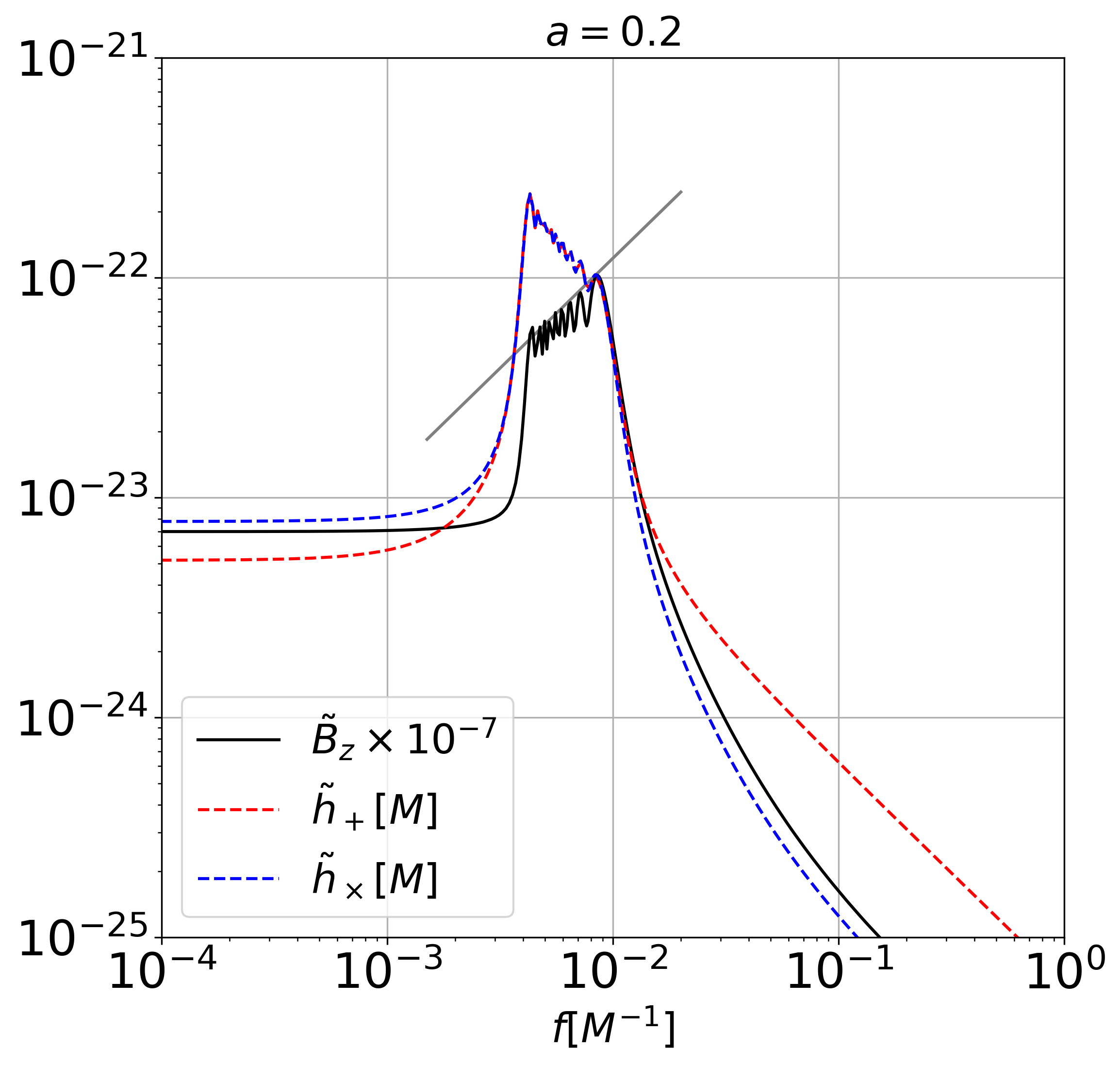}&
			\includegraphics[scale=0.36]{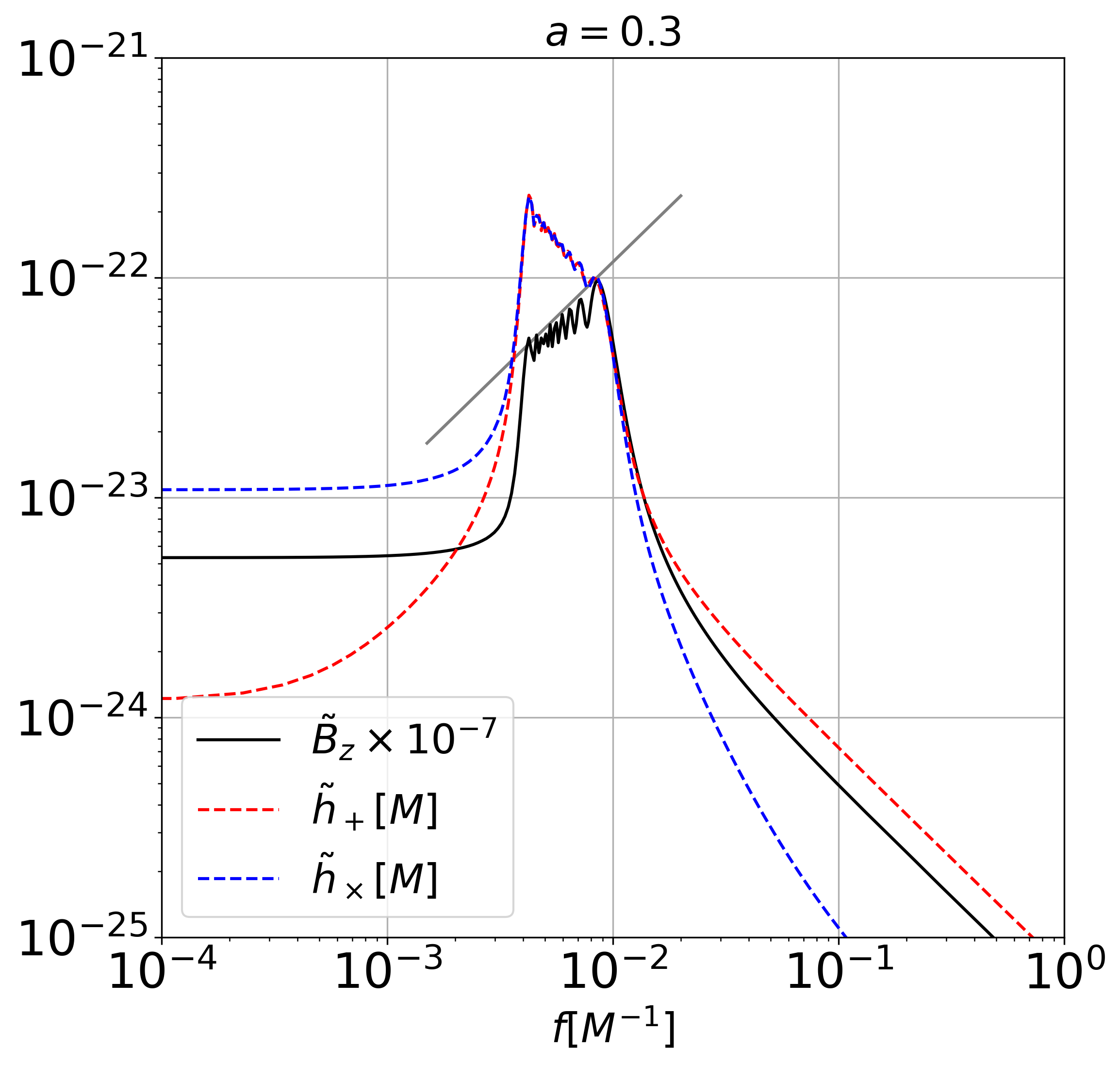}
			\end{array}$
		\end{center}
		\caption{Fourier transform of $h_+$, $h_-$ and $B_z$. Note that each value of $B_z$ should be multiplied by $10^{-7}$ in order to compare with $\tilde{h}_{+/-}$. In the figure, we also show the plot of Eq.~(\ref{4D29}), see the gray line. We use $M$ units.
		\label{fig8}}
	\end{figure*}
    
    Using Eqs.~(\ref{4Ce5}) and (\ref{4Ce10}), it is possible to obtain an analytical expression for $R$ as a function of $\tau$. Hence, taking into account that $d/d\tau=-d/dt$, one gets~\cite{Maggiore:2007ulw}
    \begin{equation}
        \label{4Ce12}
        R(t)=R_0\left(\frac{t_\text{coal}-t}{t_\text{coal}-t_0}\right)^\frac{1}{4}.
    \end{equation}
    In Eq.~(\ref{4Ce12}), $R_0$ denotes the value of $R$ at the initial time $t_0$ ($\tau_0=t_\text{coal}-t_0$).
    
    In Ref.~\cite{Maggiore:2007ulw}, M.~Maggiore also obtains an analytical expression for $\phi$ as a function of $t$ in the quasi-circular approximation. To do so, he considers the fact
    \begin{equation}
        \label{4Ce13}
        \phi(t)=\int^{t}_{t_0}\omega_s(t')dt'.
    \end{equation}
    Hence, after integration, one obtains 
    \begin{equation}
        \label{4Ce14}
        \phi(t)=-\left(\frac{5GM_c}{c^3}\right)^{-\frac{5}{8}}\tau^{\frac{5}{8}}+\phi_0,
    \end{equation}
    where $\phi_0=\phi(\tau=0)$ is an integration constant representing the value of $\phi$ at coalescence. 
  
    In Fig.~\ref{fig7}, we compare the behavior of $R(t)$ (first row) and $\phi(t)$ (second row) as functions of time. There, the values of $R$ and $\phi$ obtained via the EOB approach are shown in black color, while those values calculated using Eqs.~(\ref{4Ce12}) and (\ref{4Ce13}) are shown using the green-dashed line. 
    
    It is well-known that the gravitational field close to the binary system is strong and has significant consequences on its dynamics. Nevertheless,  Fig~\ref{fig7} shows that the quasi-circular approximation helps us to understand the behavior and evolution of the system. In this sense, from the physical point of view, it is valid to use Eqs.~(\ref{4Ce12}) and (\ref{4Ce14}) to investigate the behavior of the EM wave obtained in the last subsection. For this reason, since it is a usual procedure in signal analysis, the next step is to compute the Fourier transform of the EM wave generated by the phenomenological model.  
    

    \subsection{The Fourier transform of the EM wave\label{SecIVd}}
   
    Under certain conditions, it is well-known that a function $g(t)$ (the signal) can be expressed as a linear combination of sines and cosines using the Fourier series. At the same time, it is possible to represent the signal in the frequency domain, which can be understood as the collection of frequencies constituting $g(t)$. The representation of $g(t)$ in the frequency domain, denoted by $\tilde{g}(f)$, is known as the Fourier transform and is a helpful tool to analyze and obtain information about the evolution and behavior of the signal. For this reason, we devote this subsection to compute the Fourier transform of the EM wave. To do so, we follow once again the ideas of  Ref.~\cite{Maggiore:2007ulw}
    
    The output generated by the EOB approach is represented by discrete values. In this sense, the analysis of the GW and EM signals requires the standard fast Fourier transform (FFT), see Fig.~\ref{fig8}. Nevertheless, from the theoretical point of view, it is often much more convenient to deal with analytic expression rather than a set of discrete values. Therefore, efforts to obtain such analytical expressions, even if they are only approximate, are well justified.
    
    One can use the quasi-circular approximation to obtain analytical expressions for the GW amplitude. According to Ref.~\cite{Maggiore:2007ulw}, the behavior of $h_+$ and $h_\times$ in terms of $\tau$ is given by
    \begin{equation}
        \label{4De1}
        \begin{aligned}
            h_+(\tau)&=\frac{1}{L}\left(\frac{GM_c}{c^2}\right)^\frac{5}{4}\left(\frac{5}{c\tau}\right)^\frac{1}{4}\left(\frac{1+\cos^2\iota}{2}\right)\cos2\phi(\tau),\\
            h_\times(\tau)&=\frac{1}{L}\left(\frac{GM_c}{c^2}\right)^\frac{5}{4}\left(\frac{5}{c\tau}\right)^\frac{1}{4}\cos\iota\cos2\phi(\tau),
        \end{aligned}
    \end{equation}
    where $L=||\textbf{P}||$ is the distance from the source to the observer. Once again, it is important to recall that $\tau=t_\text{coal}-t$ is evaluated at the observer time $t$ rather than retarded time. From these equations, we can see that both the frequency and the amplitude increase as the binary system approach the coalescence. This behavior, also known as ``\textit{chirping}'', is shown clearly in the third and fourth rows of Fig.~\ref{fig3}. 
  
    Starting from Eq.~(\ref{4De1}), M. Maggiore computed the Fourier transform by means of the stationary phase method, which gives a very accurate expression for the GW signal produced by an inspiraling compact binary~\cite{Droz:1999qx}. The main idea of the method consist on the cancellation of sinusoids with rapidly varying phase. Hence, for a GW signal in the quasi-circular approximation, the Fourier transform of Eq.~(\ref{4De1}) is given by~\cite{Maggiore:2007ulw}  
    \begin{equation}
        \label{4De2}
        \begin{aligned}
        \tilde{h}_+(f)&=Ae^{\Psi_+(f)}\frac{c}{L}\left(\frac{GM_c}{c^3}\right)^\frac{5}{6}\frac{1}{f^\frac{7}{6}}\left(\frac{1+\cos^2\iota}{2}\right),\\\\
        \tilde{h}_\times(f)&=Ae^{\Psi_\times(f)}\frac{c}{L}\left(\frac{GM_c}{c^3}\right)^\frac{5}{6}\frac{1}{f^\frac{7}{6}}\cos\iota,
        \end{aligned}
    \end{equation}
    with $\iota$ the binary inclination angle, and $A$ a constant defined as~\cite{Maggiore:2007ulw}  
    \begin{equation}
        \label{4De3}
        A=\frac{1}{\pi^\frac{2}{3}}\left(\frac{5}{24}\right)^\frac{1}{2},
    \end{equation}
    and the phases are given by $\Psi(f)_\times=\Psi_++(\pi/2)$ and~\cite{Maggiore:2007ulw} 
    \begin{equation}
        \label{4De4}
        \Psi_+(f)=2\pi f\left(t_\text{coal} +\frac{L}{c}\right)-2\phi_0-\frac{\pi}{4}+\frac{3}{4}\left(\frac{GM_c}{c^3}8\pi f\right)^{-\frac{5}{3}}.
    \end{equation}
    From Eq.~(\ref{4De2}), we can see that $h_+$ and $h_\times$ are proportional to $f^{-7/6}$.
    
    In the case of the EM signal obtained using the phenomenological model, we found that the magnetic field $\textbf{B}_z$ is given by 
    \begin{equation}
    \label{4De5}
    B_z=-\frac{1}{c^3||\textbf{P}||}\left\{\mathcal{A}(t)\sin2\phi(t)+\mathcal{B}(t)\cos2\phi(t)\right\},
    \end{equation}
    where
    \begin{widetext}
    \begin{equation}
    \label{4De6}
    \begin{aligned}
    \mathcal{A}(t)&=\frac{r_H}{2}\left[3\dot{R}\left(\ddot{q}_A-4q_A \dot{\phi}^2\right)+3\dot{q}_A\ddot{R}+R\left(\dddot{q}_A-12\dot{\phi}\left(\dot{q}_A \dot{\phi}+q_A \ddot{\phi}\right)\right)+q_A \dddot{R}\right]\\\\
    \mathcal{B}(t)&=r_H\left[3R\ddot{q}_A \dot{\phi}+3 \dot{q}_A\left(2\dot{R}\dot{\phi}+R\ddot{\phi}\right)+q_A\left(3 \ddot{R} \dot{\phi}+3\dot{R}\ddot{\phi}+R\left(\dddot{\phi}-4\dot{\phi}^3\right)\right)\right].
    \end{aligned}
    \end{equation}
    \end{widetext}
    However, since the radial velocity $\dot{R}$ is neglectable as long as $\dot{\omega}_s<<\omega^2_s$ ($\omega_s\equiv\dot{\phi}$), we can approximate the value of $B_z$ given in Eq.~(\ref{4De5}) in such a way that, as long as we are in the inspiral phase, we can neglect the terms proportional to $\dot{R}$, and similar for the terms involving the derivative of $\omega_s=\dot{\phi}$~\cite{Maggiore:2007ulw}. In a similar way, because the charge $q_A$ depends on the orbital velocity $v$ (which depends on $R$, $\dot{R}$, $\phi$, and $\dot{\phi}$), the value of $\dot{q}_A$, $\ddot{q}_A$ and $\dddot{q}_A$, can be neglected. Therefore, in the quasi-circular approximation, $B_z(t)$ is given by 
    \begin{equation}
    \label{4De7}
   B_z(t)\approx\frac{4r_H R q_A\dot{\phi}^3}{c^3||\textbf{P}||}\cos2\phi(t)=-2\frac{r^3_HR^2B_0\dot{\phi}^4}{\pi c^4||\textbf{P}||}\cos2\phi(t),
    \end{equation}
    where we used the relation $v\approx R\dot{\phi}$. Now, from Eq.~(\ref{4Ce14}), we have
    \begin{equation}
        \label{4De8}
        \dot{\phi}(t)=\frac{5}{8}\left(\frac{5GM_c}{c^3}\right)^{-\frac{5}{8}}(t_\text{coal}-t)^{-\frac{3}{8}}.
    \end{equation}
    Therefore, after replacing Eqs.~(\ref{4Ce12}) and (\ref{4De8}) in Eq.~(\ref{4De7}), we obtain 
    \begin{equation}
        \label{4De9}
        B_z(t)\approx-\mathcal{D}(t)\cos2\phi(t)
    \end{equation}
    with
    \begin{equation}
        \label{4De10}
        \mathcal{D}(t)=B_0\kappa(t_\text{coal}-t)^{-1},
    \end{equation}
    and
    \begin{equation}
        \label{4De11}
        \kappa=\frac{2r^3_H}{\pi c^2||\textbf{P}||}\left(\frac{5}{256}\right)\left(\frac{GM_c}{c^3}\right)^{-\frac{5}{2}}\left(\frac{Gm}{c^3}\right)\left(\frac{G\mu}{c^3}\right)^\frac{1}{2}
    \end{equation}
    where we used the relation~\cite{Maggiore:2007ulw}
    \begin{equation}
        \label{4De11a}
        \tau_0=t_{\text{coal}}-t=\frac{5}{256}\frac{R^4_0}{c^4}\left(\frac{Gm}{c^3}\right)^{-2}\left(\frac{G\mu}{c^3}\right)^{-1}.
    \end{equation}
    
    To compute the Fourier transform of $B_z$, we need to express Eq.~(\ref{4De9}) in terms of the exponential function $e$. To do so, we use the well-known Euler relation 
    \begin{equation}
        \label{4De12}
        e^{xi}=\cos x+i\sin x,
    \end{equation}
    from which
    \begin{equation}
        \label{4De13}
        \begin{aligned}
            \cos 2\phi(t)&=\frac{e^{2i\phi(t)}+e^{-2i\phi(t)}}{2}.
        \end{aligned}
    \end{equation}
    Hence
    \begin{equation}
        \label{4D14}
        B_z(t)=-\mathcal{D}(t)\left(\frac{e^{i\Phi(t)}+e^{-i\Phi(t)}}{2}\right).
    \end{equation}
    Note that we define $\Phi(t)=2\phi(t)$ for simplicity.
    
    The Fourier transformation of $B_z(t)$ is given by
    \begin{equation}
        \label{4D15}
        \tilde{B}_z(f)=-\int dt\mathcal{D}(t_\text{r})\cos\Phi(t_\text{r})e^{2\pi fti}
    \end{equation}
    where $t_\text{r}=t-||\textbf{P}||/c$. Then $dt_\text{r}=dt$, and the last expression reduces to
    \begin{equation}
        \label{4D16}
        \begin{aligned}
        \tilde{B}_z(f)&=\frac{1}{2}e^{2\pi f\frac{||\textbf{P}||}{c}i}\big(\int d_{t_\text{r}}\mathcal{D}(t_\text{r})e^{(-\Phi(t_\text{r})+2\pi f t_\text{r})i}\\
        &-\int d_{t_\text{r}}\mathcal{D}(t_\text{r})e^{(\Phi(t_\text{r})+2\pi f t_\text{r})i}\big).
        \end{aligned}
    \end{equation}
    In the last expression, we do not write the limits of integration explicitly because the integral is computed by means of the stationary phase method. In this sense, for the moment we only need the stationary point that belongs to the integration domain $t<t_\text{coal}$. In this way, the fact that $\mathcal{D}(t)$ diverges at the limit $t=t_\text{coal}$ becomes irrelevant~\cite{Maggiore:2007ulw}. 
    
    Form Eq.~(\ref{4De8}), we see that $\dot{\Phi}=\omega_\text{GW}>0$. Therefore, only the term proportional to $e^{(-\Phi(t_\text{r})+2\pi f t_\text{r})i}$ has a stationary point, while the term proportional to $e^{(\Phi(t_\text{r})+2\pi f t_\text{r})i}$ is always oscillating fast, and integrates to a negligible small value. Hence, the Fourier transformation of the EM wave reduces to
    \begin{equation}
        \label{4D17}
        \tilde{B}_z(f)\approx \frac{1}{2}e^{2\pi f \frac{||\textbf{P}||}{c}i}\int dt_\text{r}\mathcal{D}(t_\text{r})e^{\gamma(t_\text{r})i},
    \end{equation}
    with
    \begin{equation}
        \label{4D18}
        \gamma(t_\text{r})= 2\pi f t_\text{r}-\Phi(t_\text{r}). 
    \end{equation}
    The stationary point $t_*(f)$ is determined by the condition $2\pi f= \dot{\Phi}(t_*)=\omega_{GW}$. Hence, expanding the function $\gamma$ in powers of $(t_\text{r}-t_*)$ up to second order, we obtain
    \begin{equation}
        \label{4D19}
        \begin{aligned}
        \gamma(t_\text{r})&\approx \sum^2_{n=0}\frac{\gamma^{(n)}(t_*)}{n!}(t_\text{r}-t_*)^n\\
        &=\frac{\gamma^{(0)}(t_*)}{0!}+\frac{\gamma^{(1)}(t_*)}{1!}(t_\text{r}-t_*)+\frac{\gamma^{(2)}(t_*)}{2!}(t_\text{r}-t_*)^2\\
        &=2\pi ft_*-\Phi(t_*)-\frac{\ddot{\Phi}(t_*)}{2}(t_\text{r}-t_*)^2.
    \end{aligned}
    \end{equation}
    Then, with the change of variable 
    \begin{equation}
        \label{4D20}
        \begin{aligned}
        x&=\sqrt{\frac{\ddot{\Phi}(t_*)}{2}}(t_{r}-t_*)\\
        dx&=\sqrt{\frac{\ddot{\Phi}(t_*)}{2}}dt_{r},
        \end{aligned}
    \end{equation}
    we obtain 
    \begin{equation}
        \label{4D21}
        \begin{aligned}
        \tilde{B}_z(f)&\approx \frac{1}{2}e^{2\pi f\frac{||\textbf{P}||}{c}i}\mathcal{D}(t_*)e^{\left(2\pi ft_*-\Phi(t_*)\right)i}\\
        &\times\left(\frac{2}{\ddot{\Phi}(t_*)}\right)^\frac{1}{2}\int^\infty_{-\infty}e^{-x^2i}dx.
        \end{aligned}
    \end{equation}
    Using 
    \begin{equation}
        \label{4D22}
        \int^\infty_{-\infty}e^{-x^2i}dx=\sqrt{\pi}e^{-\frac{\pi}{4}i},
    \end{equation}
    we get 
    \begin{equation}
        \label{4D23}
        \tilde{B}_z(f)\approx\frac{1}{2}e^{\Psi i}\mathcal{D}(t_*)\left(\frac{2\pi}{\ddot{\Phi}(t_*)}\right)^\frac{1}{2}
    \end{equation}
    where, we define
    \begin{equation}
        \label{4D24}
        \Psi=2\pi f\left(t_*+\frac{||\textbf{P}||}{c}\right)-\Phi(t_*)-\frac{\pi}{4}
    \end{equation}
    Now, recalling that 
    \begin{equation}
        \label{4D25}
        \tau_*(f)=\frac{5}{256}\left(\frac{GM_c}{c^3}\right)^{-\frac{5}{3}}(\pi f)^{-\frac{8}{3}}
    \end{equation}
    where $\tau_*=t_c-t_*$ and $t_c$ is the retarded time of $t_\text{coal}$. Hence we have
    \begin{equation}
        \label{4D26}
        \mathcal{D}(f)=\frac{256}{5}B_0\left(\frac{GM_c}{c^3}\right)^{\frac{5}{3}}\kappa (\pi f)^\frac{8}{3}
    \end{equation}
    On the other hand,
    \begin{equation}
        \label{4D27}
        \ddot{\Phi}(\tau)=\frac{15}{32}\left(\frac{5GM_c}{c^3}\right)^{-\frac{5}{8}}\tau^{-\frac{11}{8}}
    \end{equation}
    from which
    \begin{equation}
        \label{4D28}
        \begin{aligned}
        \left(\frac{2\pi}{\ddot{\Phi}(\tau_*)}\right)^\frac{1}{2}&=\frac{5}{4}\left(\frac{\pi}{30}\right)^\frac{1}{2}
        \left(\frac{GM_c}{c^3}\right)^{-\frac{5}{6}}(\pi f)^{-\frac{11}{6}}
        \end{aligned}
    \end{equation}
    Finally we have
    \begin{equation}
        \label{4D29}
    \begin{aligned}
    \tilde{B}_z(f)&\approx\frac{5}{32}\frac{B_0r^3_H}{\pi c^2 ||\textbf{P}||}\left(\frac{\pi}{30}\right)^\frac{1}{2}\left(\frac{Gm}{c^3}\right)^{-\frac{1}{6}}(\pi f)^\frac{5}{6}e^{\Psi i},
    \end{aligned}
    \end{equation}
    where we had into account the relations $\mu=m/4$, and  $M_c=m/\sqrt[6]{2}$, see App.\ref{A1}. In fig.~\ref{fig8}, we show the plot of Eq.~(\ref{4D29}), see the gray line. 
    
	
	\section{Discussion and conclusion \label{SecV}}
	
    According to Ref.~\cite{Palenzuela:2009yr}, except for systems formed solely by black holes, the majority of astrophysical systems which produce gravitational waves may also generate EM radiation. Hence, different mechanisms could be crucial in creating EM waves during the coalescence of compact binary systems~\cite{Sylvestre:2003vc}. For example, when a neutron star (NS) orbits a companion with a strong magnetic field, an electric field is induced in the orbiting star. Hence, this interaction leads to particle acceleration generating a stellar wind and coherent EM radiation like normal pulsars. A second mechanism is the radioactive decay of the neutron-rich nuclei of the decompressed NS matter ejected during the merger. This ejection can produce the energy required to power an EM signal. Finally, another mechanism appears from the possibility that the merger of a compact binary system can generate a relativistic blast wave, which could explain the long-wavelength counterparts to gamma-ray bursts.
    
    Undoubtedly, the strong and highly dynamical gravitational fields around compact objects can affect the dynamics of plasmas and matter~\cite{Palenzuela:2009hx}. In this sense, the scenario of a single black hole interacting with an accretion disk is an excellent example of such a system. Thanks to the works of Penrose~\cite{Penrose:1969pc}, and Blandford and Znajek~\cite{Blandford:1977ds}, this scenario is well understood, and possible mechanisms to explain the highly energetic emissions from single black hole systems interacting with surrounding plasma has been proposed~\cite{Lee:1999se}. Nevertheless, in the context of galaxy mergers, strong EM emissions could occur before the scenario of a pseudostationary, single black hole interacting with an accretion disk, i. e. when individual black holes in each galaxy eventually collide in the galaxy resulting from the merger.
    
    The GW signal produced by the collision of two black holes at the center of each galaxy would be detected by laser interferometric space antenna (LISA). Furthermore, according to Refs.~\cite{Milosavljevic:2004cg, Haiman:2009te}, it is known that way before the merger, the binary system ``hollows out" any surrounding gas. In the beginning, the hollow follows the shrinking binary inward. But eventually, the gravitational radiation timescale becomes shorter than the viscous timescale in the disk~\cite{Milosavljevic:2004cg}. Therefore, the final state is a merged black hole surrounded by a circumbinary disk.
    
    Different works have investigated the pseudostationary stage mentioned above, showing that electromagnetic counterparts can be produced~\cite{Palenzuela:2009yr, Palenzuela:2009hx, Chang:2009rx,vanMeter:2009gu, Kelly:2017xck, Kelly:2020vpv}. It is well-known that modeling such electromagnetic emissions requires investigating the behavior of the gas and the fields in the strong-field regions that surround the binary system. Nevertheless, to understand the process of EM emission, Palenzuela et al. have suggested a simple model for this process long before the merger, the in-spiral. In Refs.~\cite{Palenzuela:2009yr, Palenzuela:2009hx}, the authors consider the collision of two black holes surrounded by a circumbinary disk, which has anchored a magnetic field. As claimed by the authors, a good idealization well before the merger is to consider a set of point charges that orbit circularly. In this sense, in our work, we focused our analysis only on the in-spiral stage, where the quasi-circular approximation can be considered. Since the dominant features of the GW wave from the inspiral phase are captured by neglecting the spins and internal structure of the binary elements~\cite{Bartos:2012vd}, we used the Newtonian approximation to model the evolution of $R(t)$ and $\phi(t)$ to obtain the Fourier transform. 
    
    The amplitude of the magnetic field increases as the two black holes approach each other. 
    Here, it is important to point out that the magnitude of the EM field shown in the figures only considers the contribution due to the motion of the charges. The total magnetic field takes into account the magnetic field anchored to the circumbinary disk, i. e. $\textbf{B}_\text{Total}=\textbf{B}_0+\textbf{B}_\text{distribution}$. In this sense, the magnetic field generated by the charge distribution has an order of magnitude of $10^{-13} T$ ($10^{-9}\text{Gauss}$) while the order of magnitude of $h_+$ and $h_\times$ is $10^{-21}$ to an observer at 1 Gly. This behavior is expected due to the form of the EM field at large distances from the source, where the dipole contribution on the EM field vanishes, and the quadrupole contribution, although small, becomes important.
    
    Our model only considers a uniform magnetic field that is perpendicular to the orbital plane. Any consideration regarding the properties of the plasma around the binary system has not been taken into account. In this sense, the EM wave can propagate freely in space once produced. However, in a realistic astrophysical scenario, the presence of plasma will affect the propagation of the EM signal. In this situation, the binary system is completely enveloped by a region of gas, which would block any radiation or other outflows. Even in more detailed works regarding the EM counterpart from BH-BH binary systems, the large temporal and spatial scales needed to simulate the EM radiation are not appropriate for an observer located at far distances~\cite{Kelly:2017xck}. For this reason, most of the works focus the analysis on near-zone mechanisms that could drive EM outflows.
    
    One of the most interesting features obtained by this analysis comes from the Fourier transform of the EM signal. In the case of GWs, we know that they can be resolved into two linearly polarised components $h_+$ and $h_\times$. The effect of GWs on a ring of test particles depends on the polarization. In the case of $h_+$, the ring deforms into an ellipse that pulsates in and out in the $x$ and $y$ directions. The ``\textit{cross}'' polarization, on the other hand, has the same effect but at an angle of $45\degree$. When the Fourier transform of $h_+$ and $h_\times$ is computed, we see that each polarization is proportional to $f^{-7/6}$ (see Eqs.~(\ref{4De2})). In this sense, when we plot the FFT obtained from the data, it is possible to see how its value decreases as $f$ increases. Nevertheless, in contrast to the GW wave, when we compute the Fourier transform of the electromagnetic signal (the contribution from the motion of the charges in the distribution), we note a different behavior: the Fourier transform of the magnetic field is proportional to $f^{5/6}$, see Eqs.~(\ref{4D29}). Therefore, the FFT increases its value as the frequency increases its value.
    
    These results show that the behavior of the EM waves is similar to that of the GWs. The frequency of EM waves is the same as the one of GWs, and the EM waves also demonstrate a kind of chirp characteristic, though the power-law versus frequency is 5/6 instead of -7/6 for GWs. Because of these properties, we may call these kinds of EM waves from binary BHs as the EM response of GWs. 
    
    The possibility of detecting this EM response depends on the frequency and strength. For the stellar-mass BHs, the EM waves frequency is from a few hundred to a few thousand Hz, which may not propagate in the Universe due to the cutoff frequency ($\gtrsim$ 500 Hz \cite{Ferriere:2001, Rybichi:2004}) of the interstellar medium. Fortunately, the merger of two neutron stars with a strong magnetic field may produce this kind of EM response with frequency 3000 - 4000 Hz, which can propagate through the Universe and finally arrive at our Solar system. In addition, primordial black holes with subsolar mass will radiate EM response with higher frequency. As an example, based on the parameters in this paper, the varied magnetic field at the solar system is about $10^{-9}$ Gauss from a source at 1 Gly. In principle, this strength can be detected using some sensitive magnetometers like superconducting quantum interference devices (SQUID). In these senses, the detection of such kinds of EM waves may be still considerable. However, in the present paper, we do not want to discuss this point in deep.  
    
    Finally, we note the possibility of extending this model to a binary system formed by charged black holes\footnote{This work is already in process and we hope to submit it in a new manuscript}. As mentioned before in the introduction, recently, L. Liu et al. have considered the gravitational and electromagnetic radiation of BH binaries, with electric and magnetic charges. In this sense, we can use similar ideas to obtain the waveform profile generated by such binary systems. This work is in progress, and we expect to present it in a future manuscript.

	
	\begin{acknowledgments}
	The work of C.A.B.G. is supported by a postdoc fund through PIFI of the Chinese Academy of Sciences. W. B. H. is supported by CAS Project for Young Scientists in Basic Research YSBR-006, and NSFC No. 11773059, 12173071, 12111530107. C.A.B.G also acknowledges the help of Xingyu Zhong for the EOB data, and Prof. Zhoujian Cao for helpful discussion. 
	\end{acknowledgments}

	\newpage
	\appendix
	\section{ $M$-units \label{A1}}
	
	As mentioned before, we use $M$-units for all the plots in this manuscript. In this sense, for clearness, we devote this appendix to discuss the process of expressing the variables in $M$-units.
	
	In GR, it is common to use geometrized units. In this system of units, $G=c=1$, and all quantities are expressible in the power of length. 
    One example of geometrized units is the $M$-units. In this case, one represents all quantities using powers of $M$, which can be the mass of the Sun or the total mass of a system. In geometrized units, the mass is given by
    \begin{equation}
        \label{A1e1}
        M=\frac{Gm}{c^2}.
    \end{equation}
    Here $G=6.67408\times 10^{-11}\text{m}^{3}\text{kg}^{-1}\text{s}^{-2}$ is the Newton constant, $c=299792458 \text{m}\text{s}^{-1}$ the speed of light and $m$ the mass in kg . In our case, we set the value of $m$ as the total mass of the binary system, i. e. $m=m_1+m_2$, where $m_1$ and $m_2$ denote the mass of each black hole. The data from the EOB simulation was obtained using $m_1=m_2=10M_\odot$, with $M_{\odot}=1.989\times 10^{30}\text{kg}$ the mass of the Sun. Therefore, we have
    \begin{equation}
        \label{A1e2}
        \begin{aligned}
            M&=\frac{G(20M_{\odot})}{c^2}\\&=\frac{(6.673\times 10^{-8}\text{cm}^3\text{g}^{-1}\text{s}^{-2})(20\times 1.989\times 10^{33}\text{g})}{(2.997 924 562\times 10^{10} \text{cm}\text{s}^{-1})^2}\\
        &\approx 2.95413\times 10^6\text{cm}.
        \end{aligned}
    \end{equation}
    This means that $1M$ is equivalent to $29540.292\text{m}$. Using this equivalence, we can express the observer distance $L=||\textbf{P}||=1\times10^9\text{ly}$ (the distances at which the GW150914 event took place) in $M$-units. Hence, from Eq.~(\ref{A1e2}), we obtain 
    \begin{equation}
        \label{A1e3}
        \begin{aligned}
           L&=1\times10^9\text{ly}\times\frac{9.461\times10^{12}\text{km}}{1\text{ly}}\times\frac{10^5\text{cm}}{1\text{km}}\times\frac{1M}{2.95413\times 10^6\text{cm}}\\
           &\approx 3.20265\times 10^{20}M. 
        \end{aligned}
    \end{equation}
    Similar to the mass, we can express the time in $M$-units. In GR, the time in geometrized units is given by 
    \begin{equation}
        \label{A1e4}
        \begin{aligned}
            M&=\frac{Gm}{c^3}=\frac{G(20M_{\odot})}{c^3}\\
            &=\frac{G(20M_{\odot})}{c^2}\\&=\frac{(6.673\times 10^{-8}\text{cm}^3\text{g}^{-1}\text{s}^{-2})(20\times 1.989\times 10^{33}\text{g})}{(2.997 924 562\times 10^{10} \text{cm}\text{s}^{-1})^3}\\
            &\approx 9.85391\times10^{-5}\text{s}.
        \end{aligned}
    \end{equation}
    Hence, to convert from $\text{s}$ to $M$, we do the following operation 
    \begin{equation}
        \label{A1e5}
        t[M]=\frac{t[\text{s}]}{9.85391\times10^{-5}\text{s}}M.
    \end{equation}
    
   From Eq.~(\ref{4Ce1}), the \textit{reduced mass} $\mu$ in terms of $m=m_1+m2$ is given by
   \begin{equation}
       \label{A1e6}
       \begin{aligned}
          \mu&=\frac{m_1m_2}{m_1+m_2}=\frac{100M^2_\odot}{20M_{\odot}}=5M_\odot=\frac{5}{20}(20M_\odot)=\frac{m}{4},
       \end{aligned}
   \end{equation}
   and from Eq.~(\ref{4Ce1}), the \textit{chirp mass} $M_\text{c}$ in terms of $m$ is 
   \begin{equation}
       \label{A1e7}
       \begin{aligned}
       M_c&=\frac{(m_1m_2)^\frac{3}{5}}{(m_1+m_2)^\frac{1}{2}}=\frac{(100M_\odot^2)^\frac{3}{5}}{(20M_\odot)^\frac{1}{5}}=\frac{m}{\sqrt[6]{2}}.
       \end{aligned}
   \end{equation}


	\end{document}